\documentclass[reprint, amsmath,amssymb, aps, pre
]{revtex4-2}

\usepackage[utf8]{inputenc}
\usepackage{graphicx}% Include figure files
\usepackage{dcolumn}% Align table columns on decimal point
\usepackage{bm}% bold math
\usepackage[font=small, justification=justified, singlelinecheck=false]{caption}
\usepackage{subcaption}
\usepackage{hyperref}% add hypertext capabilities
\usepackage{tcolorbox}
\usepackage[dvipsnames]{xcolor}

\definecolor{Myorange}{cmyk}{0,0.5,1,0}

\newcommand{\corr}[1]{\textcolor{black}{#1}}%\textbf{#1}}} blue
\newcommand{\corrb}[1]{\textcolor{black}{#1}}%\textbf{#1}}} red

\begin{document}
%\preprint{APS/123-QED}
\title{Hyperbolic embedding of multilayer networks}% Force line breaks with \\

\author{Martin Guillemaud}
\email{martin.guillemaud@gmail.com}
%\altaffiliation{These authors contributed equally to this work}
\affiliation{Paris Brain Institute (ICM), CNRS UMR7225, Inserm U1127, Sorbonne University UM75, Inria-Paris. Pitié Salpêtrière University Hospital. Paris, France}

\author{Vera Dinkelacker}
\affiliation{Department of Neurology, Strasbourg University Hospital, Strasbourg, France}

\author{Mario Chavez}
\affiliation{CNRS, Pitié Salpêtrière University Hospital. Paris, France\\}

\date{\today}% It is always \today, today,

\begin{abstract}
Multilayer networks offer a powerful framework for modeling complex systems across diverse domains, effectively capturing multiple types of connections and interdependent subsystems commonly found in real-world scenarios. To analyze these networks, embedding techniques that project nodes into a lower-dimensional geometric space are essential. This paper introduces a novel hyperbolic embedding framework that advances the state of the art in multilayer network analysis. Our method, which supports heterogeneous node sets across networks and inter-layer connections, generates layer-specific hyperbolic embeddings, enabling detailed intra-layer analysis and inter-layer comparisons, while simultaneously preserving the global multilayer structure within hyperbolic space —a capability that sets it apart from existing approaches, which typically rely on independent embedding of layers. Through experiments on synthetic multilayer stochastic block models, we demonstrate that our approach effectively preserves community structure, even when layers consist of different node sets. When applied to real brain networks, the method successfully clusters disease-related brain regions from different patients, outperforming layer-independent approaches and highlighting its relevance for comparative analysis. Overall, this work provides a robust tool for multilayer network analysis, enhancing interpretability and offering new insights into the structure and function of complex systems.
\end{abstract}

%\keywords{Suggested keywords}%Use showkeys class option if keyword display desired
\maketitle

\section{Introduction}\label{Sec:intro}
The network paradigm has proven to be a powerful and versatile framework for representing and analyzing complex systems across a wide range of domains, from social dynamics and economic systems to neuroscience and biological interactions~\cite{Boccaletti2006}. In many real-world cases, however, interactions within a system are not adequately captured by a single-layer network, as multiple types of connections or interdependent subsystems often coexist. Multilayer network models were introduced to overcome these limitations, offering a comprehensive and structured way to represent complex systems with multiple dimensions of interaction~\cite{Boccaletti2014, Kivel2014}. In this framework, each layer typically corresponds to a specific mode of interaction (or a temporal snapshot in a time-varying model), and nodes can participate in one or several layers. Among the various types of multilayer networks, multiplex networks (where the same set of nodes interact through different types of links across layers) constitute a particularly common subclass. Multilayer models have shown great utility in diverse applications, including the study of brain connectivity, community detection, and the dynamics of spreading processes~\cite{Xu2020, Puxeddu2021, Boccaletti2014}.

In recent years, graph embedding techniques have emerged as an essential tool for simplifying and analyzing large-scale networked data. The core idea of graph embedding is to project nodes into a lower-dimensional geometric space such that their proximity reflects structural similarity in the original network~\cite{Goyal2018, Cui2019}. By representing networks as vectors or functions, a wide variety of machine learning algorithms can be used to provide a solution for various downstream tasks such as network visualization, link prediction, node classification and node clustering~\cite{xu2021understanding}. A broad array of embedding techniques has been developed for single-layer networks, relying on matrix factorization, random walks, or machine learning approaches~\cite{Goyal2018, Cui2019}.

Extending these embedding frameworks to multilayer networks has attracted growing interest, leading to a variety of methods that seek to preserve both intra- and inter-layer structural information. These include spectral methods~\cite{Liu2020}, random walk-based approaches~\cite{PioLopez2021, Liu2017}, and information-theory based techniques such as mutual information maximization across layers~\cite{Wang2023, Park2020}. Other strategies use deep learning frameworks (e.g., autoencoders) to generate embeddings useful for multilayer clustering~\cite{Wang2023b}, or apply tensor factorization techniques for network decomposition and analysis, particularly in neuroscience~\cite{Liu2018}. However, most of these methods ultimately produce a single unified embedding for the entire multilayer network. While such summarization is often effective, it may overlook important layer-specific features. In scenarios where layer-level differences are meaningful (such as in inter-layer influence quantification, or comparative studies of multiple networks) having one embedding per layer, while still accounting for multilayer dependencies, is crucial. To our knowledge, few methods explicitly address this need. One such exception is the optimization-based approach of~\cite{Li2018}, which constructs layer-specific embeddings that reflect both within-layer and cross-layer dynamics.

Hyperbolic geometry has emerged as a promising framework for graph representation, as it naturally aligns with the structural characteristics of many real-world networks, including hierarchical organization, scale-free degree distributions, and high clustering coefficients~\cite{Bog2008, Krioukov2010, Bogu2021}. In tree-like structures, for example, the number of nodes increases exponentially with distance from high-hierarchy regions, necessitating dimensionalities that exceed the representational limits of Euclidean space. Moreover, increasing the dimensionality in Euclidean embeddings results in greater computational complexity and significant distortion~\cite{nickel2017poincare}. These challenges have driven the development of models that embed networks in hyperbolic space, such as PSO~\cite{Papadopoulos2012}, HyperMap~\cite{Papadopoulos2015}, Mercator~\cite{Garcia2019}, and its high-dimensional extension, D-Mercator~\cite{Jankowski2023}. In parallel, machine learning-based approaches like Coalescent embedding~\cite{Muscoloni2017} and its extensions~\cite{Kovcs2021} have provided scalable, data-driven alternatives for hyperbolic network embedding.

Despite these developments, the application of hyperbolic embedding to multilayer networks remains largely unexplored. A common strategy involves aggregating all layers into a single connectivity matrix (typically by summing or averaging the adjacency matrices) before applying conventional hyperbolic embedding methods. However, this simplification obscures the inherent multilayer structure and fails to capture the complex interactions across different types of connections. Temporal networks, often represented as multilayer graphs with each layer corresponding to a distinct time step, have received comparatively greater attention. For instance, in Ref.\cite{Wang2021}, a method was proposed for embedding temporal graphs into the hyperboloid model using unsupervised random walks. In Ref.\cite{Yang2023}, an alternative approach was introduced, embedding temporal graphs into the Poincaré ball through a memory mechanism and tangent space optimization. Notably, in Ref.~\cite{Kleineberg2016}, it was demonstrated that embedding each layer of a multilayer network independently into the hyperbolic disk, while analyzing them collectively, improves performance in tasks such as community detection, link prediction, and navigability. These findings underscore the importance of preserving layer-specific information within hyperbolic representations to fully exploit the structural complexity of multilayer networks.

In this work, we introduce a novel multilayer embedding framework in hyperbolic space that addresses these limitations. Our method produces a separate embedding for each layer while accounting for the global multilayer structure, enabling both intra-layer analysis and inter-layer comparison. This approach opens new avenues for understanding complex systems through the lens of hyperbolic geometry, with applications ranging from brain networks to social systems.

\section{Methods}\label{Sec:methods}
% Intro 
Hyperbolic geometry concerns spaces characterized by constant negative curvature $K$, which deviate fundamentally from the principles of Euclidean geometry. In hyperbolic embedding, a graph is first mapped onto a hyperboloid and can subsequently be projected onto a two-dimensional model of hyperbolic space, such as the Poincaré disk or the Klein disk model. The Poincaré disk is the unit disk $\mathbb{D} = \{ x \in \mathbb{R}^2 : \|x\| < 1 \}$ equipped with the Riemannian metric:
\begin{equation}
    ds^2 = \frac{4 \left\| dx \right\|^2}{\left(1 - \left\| x \right\|^2 \right)^2} \, ,
\end{equation}
which induces the following distance function between two points $x, y \in \mathbb{D}$:
\begin{equation}
    d_{\mathbb{D}}(x, y) = \operatorname{arcosh} \left( 1 + 
    \frac{2 \left\| x - y \right\|^2}{\left(1 - \left\| x \right\|^2 \right)
    \left(1 - \left\| y \right\|^2 \right)} \right) \, .
    \label{eq_hyp_dist}
\end{equation}

Our methodology extends the Coalescent embedding framework, initially developed for mapping single-layer networks onto the hyperbolic disk~\cite{Muscoloni2017}. In this study, we introduce two principal advancements to the original approach. First, we generalize the method to support multilayer network structures, enabling more comprehensive modeling of complex systems. Second, we adapt the embedding scheme to project networks into the Poincaré disk $\mathbb{D}$, thereby exploiting its geometric characteristics for enhanced representational accuracy.

% Preweighing
\subsection{Pre-weighting}
\label{sec_pre_weight}
The first step of our multilayer embedding is edge pre-weighting. If edge weights already reflect importance, this step can be skipped. Pre-weighting assigns new weights based on local topological importance using a ``repulsion-attraction rule''. For an edge $e_{i,j}$ between nodes $i$ and $j$, the new weight is defined as follows:
\begin{equation}
    W_{\mathrm{RA}}\left(e_{i,j}\right) = 
    \frac{d_i + d_j + d_i d_j}{1 + CN_{ij}} \, ,
\end{equation}
where $d_i$ is the degree of the node $i$ and $CN_{ij}$ the number of common neighbors between nodes $i$ and $j$. 

As in the original methodology~\cite{Muscoloni2017}, edge betweenness centrality can also be used as an alternative weighting strategy during the pre-weighting step. Nevertheless, nodes that were not previously connected remain unconnected. The framework is applicable to all edges in a multilayer network, including interlayers edges when present. 

% Global connectivity matrix
\subsection{Global Connectivity Matrix Construction – Case of multilayer Network with a Constant Node Set Across Layers}
Once the pre-weighted step has been completed or skipped, a new global connectivity matrix is constructed, incorporating the data from all layers. For clarity, we will denote $n_l$ the number of nodes present in the layer $l$ and let $L$ represents the total number of layers in the network.
We begin by considering the case where all layers share the same set of nodes, i.e., $n_l = n$ for all $l$. We then discuss the adaptations required when layers contain different node sets.

\begin{figure}[h!]
    \includegraphics[width=\columnwidth]{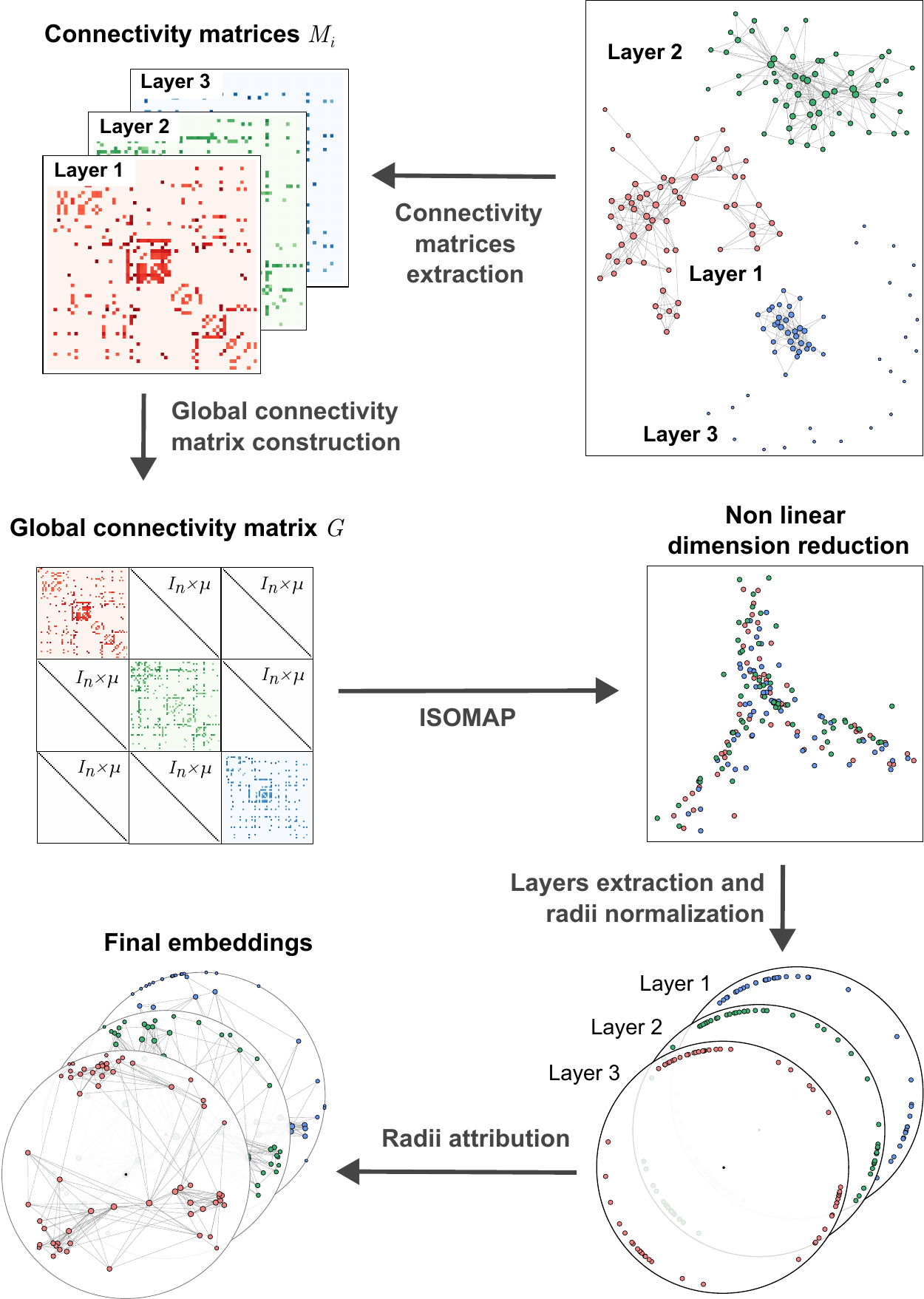}
    \caption{Global pipeline for multiple network embedding with identical nodes set across layers. The process begins with the extraction of the connectivity matrices for each layer. These matrices are then combined to form a global connectivity matrix $G$. The dimension reduction algorithm is applied to $G$ to obtain a two-dimensional representation of the dataset. From this embedding, the angular coordinates of the nodes in each layer are extracted, and their radii are initially normalized to one. In a final step, the radial coordinates are reassigned based on the centrality of each node.
 }\label{mat_cons}
\end{figure}

As illustrated in Fig.~\ref{mat_cons}, the matrix $G$ is constructed as follows.  Let us denote the connectivity matrix of layer $i$ as $M_{i,j}$ with dimensions of $n \times n$. The global connectivity matrix of the whole network will be designated as $G$, with dimensions of $n \times L$ for rows and columns~\cite{Boccaletti2014}. The matrix is divided into $L \times L$ blocks of size $n \times n$ each, noted $g_{i,j}$. The blocks on the diagonal will be occupied with the connectivity matrices of the layers, that is to say: 
\begin{equation}
    g_{i,i} = M_{i,i} \ \text{for} \ i \in \{ 1 , \dots, L \}
\end{equation}

%-------------Figure ------------

The off-diagonal matrices represent the coupling terms between the layers. In the case of disconnected layers, each off-diagonal matrix is composed of an identity matrix of size $n \times n$, scaled by a coupling coefficient $\mu$, which represents the coupling interaction between layers $i$ and $j$: 
\begin{equation}
    g_{i,j} = I_n \cdot \mu, \ \text{for}\  i,j \in \{ 1 , \dots, L \}, \; i \neq j,
\end{equation}
where $g_{i,j}$ represents the block matrix of size $n \times n$ in $G$ located in row $i$ and column $j$. Here, $I_n$ denotes the identity matrix of size $n \times n$. 

In the case where edges exist between layers (e.g., $M_{i,j}[l,m] = 1$ if an edge between node $l$ of layer $i$ is connected to node $m$ of layer $j$, preweighting step can also be applied to this connections according to Sec.~\ref{sec_pre_weight}), the off-diagonal block matrices representing the coupling between the layers will take the following form:
\begin{equation}
    g_{i,j} = I_n \cdot \mu + M_{i,j}, \ \text{for } i,j \in \{ 1 , \dots, L \}, \; i \neq j, 
    \label{eq_off_diagonal}    
\end{equation}
where $M_{i,j}$ is the interaction matrix between nodes of layers $i$ and $j$.

% CASE NOT ALL THE SAME NUMBE OF NODES ?
\subsection{Extension to multilayer graphs with heterogeneous node sets}
The method introduced in the previous subsection is formulated for multilayer graphs in which all layers share an identical set of nodes. Nevertheless, in practical applications, it is common to encounter layers comprising distinct node sets. The proposed approach remains applicable in such scenarios, with only minor modifications, as described below.

Let $n_i$ denote the number of nodes in layer $i$. To construct the global connectivity matrix $G$, the off-diagonal blocks (originally identity matrices scaled by coupling parameters $\mu$) are replaced with rectangular matrices when node counts differ across layers. Specifically, the off-diagonal block $g_{i,j}$ (for $i \neq j$) is of size $n_i \times n_j$ and contains entries equal to $\mu$ at positions corresponding to interlayer node associations. That is, if node $a$ in layer $i$ corresponds to node $b$ in layer $j$  and there is no direct link between them, then the entry $(a,b)$ in $g_{i,j}$ is simply assigned the value $\mu$, otherwise it is set to zero. In the absence of correspondence between layers nodes, the corresponding values are assigned a value of zero. When inter-layer edges (regardless of whether they are weighted) are explicitly included in the multilayer graph, they are likewise represented in the off-diagonal blocks of $G$, as specified in Eq.~\ref{eq_off_diagonal}, irrespective of whether the node sets across layers are identical. The construction of the diagonal matrices follows the same methodology employed in the standard multiplex framework. 

% Embedding method  ISOMAP 
\subsection{Dimension reduction}
Following the original approach proposed in Ref.~\cite{Muscoloni2017}, once the global connectivity matrix $G$ has been constructed, a non-linear dimensionality reduction algorithm, such as Isomap or Laplacian eigenmaps~\cite{von2007tutorial}, is applied to embed the network into a two-dimensional space. The matrix $G$ serves as input to the algorithm, which produces 2D coordinates for all $\sum_{l=1}^L n_l$ nodes, where, for instance, the positions of the first $n_1$ nodes correspond to those in layer $1$. For illustration, see Fig.~\ref{mat_cons} in the case where $n_l = n$ for all $ l$. In the present study, we adopt the Isomap algorithm, as it performs a global optimization that preserves the overall network structure by incorporating geodesic (i.e., shortest path) distances between nodes~\cite{Tenenbaum2000}.
%% Layers extraction 
%% Radius attribution 
\subsection{Layers extraction and radii attribution}
Once the non-linear dimensionality reduction has been performed, nodes retain their angular coordinates in the two-dimensional embedding, while their radial positions are reassigned based on their centrality within the network. As illustrated in Fig.~\ref{mat_cons}, the initial step involves extracting the positions of the nodes for each layer and setting all radial coordinates to $1$.  Radial positions are then reassigned according to each node’s centrality, measured by its degree. To obtain a more informative estimate of centrality, we use the weighted degree $w_i$, which accounts for the strength of connections rather than merely their number.\corr{The weighted degree of a node is defined as the sum of the weights of its incident edges, providing a measure of the node’s overall connectivity strength within the weighted network.} Notably, this centrality measure reflects any pre-weighting applied to the edges; in the absence of such preprocessing, the original edge weights are used. Although weighted degree is employed here, alternative centrality metrics may also be used, depending on the specific application or network characteristics\cite{Muscoloni2017, battiston2014structural}. Given a node $i$ with (weighted) degree $w_i$, its radial coordinate is updated as follows:
\begin{equation}
    r_i = 1 - \tanh\left( \frac{w_i}{\beta} \right) \, .
\end{equation}
where $r_i$ denotes the radial coordinate of node $i$, $\tanh$ is the hyperbolic tangent function, and $\beta$ is a scaling parameter. This formulation ensures that all nodes are embedded within the Poincaré disk, with nodes of degree zero mapped at the boundary ($r = 1$) and the most central nodes positioned closer to the origin. The parameter $\beta$ controls the rate at which the radius decreases with increasing centrality and plays a role analogous to the popularity fading parameter in the original coalescent embedding method.\corrb{Unlike the original coalescent embedding, which assigns radii based on the degree ranking within a layer, here we directly use the (weighted) degree itself. This choice ensures that nodes with the same degree across different layers are mapped at consistent radial distances, while also naturally placing disconnected nodes ($w_i=0$) at the boundary ($r=1$). This adaptation makes the method more suitable for multilayer networks, where degree distributions can differ significantly between layers.} \corr{A practical strategy for setting $\beta$ consists in prescribing the radius $r_i$ of a node $i$ with maximal (weighted) degree $w_i$, such that it lies within a desired distance from the origin (for instance, $r_i = 0.1/0.2$) and solving $\beta = w_i / \text{arctanh}(r_i)$. This ensures that node radii span a broad portion of the disk while maintaining an interpretable mapping between centrality and geometry.}

\subsection{Gaussian distributions in the Poincaré disk}

\label{sec_gaussian_dis}

The proposed embedding method exhibits the ability to consistently cluster critical nodes across multiple layers. To quantitatively assess this property, we introduce a Gaussian model in hyperbolic space to characterize the distribution of a given node's embedded positions across layers. The underlying rationale is that the covariance among the projections of the same node from different layers decreases as these projections converge toward similar coordinates. Consequently, a Gaussian model offers an appropriate statistical framework for assessing the localization a node's position across layers.

To define the parameters of the hyperbolic Gaussian distribution (specifically the barycenter and covariance) we employ the Klein model of hyperbolic space. Like the Poincaré model, the Klein model is defined over the open unit disk $\mathbb{D}$, that is , $\mathbb{K} =\{ x \in \mathbb{R}^2 : \|x\| < 1 \}$, but it is endowed with a different metric $\left( ds^2 = \frac{\left\| dx \right\|^2}{1 - \left\| x \right\|^2 } + \frac{\left( x \cdot dx \right)^2}{\left(1 - \left\| x \right\|^2 \right)^2} \right)$. Consequently, a given point will generally have different coordinate representations in the Poincaré and Klein models, due to the distinct metrics underlying each. 

Let $X_{\mathbb{D}}$ and $X_{\mathbb{K}}$ denote the coordinates of the same point in the Poincaré and Klein disk models, respectively. The bijective mapping between these representations is defined by:
\begin{equation}
    X_{\mathbb{D}} = \frac{X_{\mathbb{K}}}{1 + \sqrt{1 - \left\| X_{\mathbb{K}} \right\|^2}} \, , 
    \quad 
    X_{\mathbb{K}} = \frac{2 X_{\mathbb{D}}}{1 + \left\| X_{\mathbb{D}} \right\|^2} \, 
    \label{KleinPoinBij}
\end{equation}

Unlike other models of hyperbolic geometry, the Klein model admits a particularly simple expression for the hyperbolic barycenter (a generalization of the Euclidean mean) in Klein coordinates~\cite{Khrulkov2020}:
\begin{equation}
    \operatorname{HypBary}\left(X_1, \dots, X_n \right) = 
    \frac{\sum_{i=1}^{n} \gamma_i X_i}{\sum_{i=1}^{n} \gamma_i}, 
\end{equation}
where $  \gamma_i = \frac{1}{\sqrt{1 - \left\| X_i \right\|^2}} \,$ is the Lorentz factor associated with $X_i$. Once the barycenter is obtained in the Klein disk, it is mapped back to the Poincaré disk using Eq.~\ref{KleinPoinBij}. 

To estimate the hyperbolic Gaussian distributions in the Poincaré Disk representation, we project the node positions onto the tangent space at the barycenter —an Euclidean space where Gaussian statistics are well-defined. Specifically, the logarithmic map $\mathrm{Log}_z(X)$ projects a point $X \in \mathbb{D}$ onto the tangent space $T_z\mathbb{D}$ centered at $z \in \mathbb{D}$, while the exponential map $\mathrm{Exp}_z(V)$ performs the inverse operation, mapping a tangent vector $V \in T_z\mathbb{D}$ back to the hyperbolic disk. These maps are defined as follows:
\begin{equation}
    \operatorname{Log}_z\left(X\right) = \frac{2}{\lambda_z} \operatorname{arctanh}\left( \left\| -z \oplus X \right\| \right)
    \frac{-z \oplus X}{\left\| -z \oplus X \right\|} \, ,
\end{equation}

\begin{equation}
    \operatorname{Exp}_z(V) = z \oplus \left( \tanh\left( \frac{\lambda_z \left\| V \right\|}{2} \right)
    \frac{V}{\left\| V \right\|} \right) \, 
\end{equation}

Here, $\oplus$ denotes the Möbius addition, defined by~\cite{ungar2001hyperbolic}:
\begin{equation}
    X \oplus Y = \frac{ \left(1 + 2 \langle X, Y \rangle + \left\| Y \right\|^2 \right) X
    + \left( 1 - \left\| X \right\|^2 \right) Y }{1 + 2 \langle X, Y \rangle + \left\| X \right\|^2 \left\| Y \right\|^2} \,
\end{equation}

To estimate the covariance matrix of a set of points $X_i$ in the hyperbolic disk, we first compute their barycenter $z$ in the Klein disk and subsequently map it back to the Poincaré disk. The set of points is then projected onto the tangent space $T_z\mathbb{D}$, centered on the estimated barycenter $z$ using the logarithmic map $\mathrm{Log}_{z}(\cdot)$. Within this Euclidean subspace, the barycenter is the point with coordinates $(0, 0)$ and the covariance matrix can  be calculated as: $V_{z} = \mathrm{Covariance} \left( \left[ \mathrm{Log}_{z}(X_i) \right]^T \right)$.

For a given position $\tilde{X}$ in the Poincaré Disk, the hyperbolic Gaussian pdf of the nodes $X_i$ estimation is : 
\begin{equation}
       \mathrm{Pdf}_z\left( {\tilde{X}}  \right) =  \frac{1}{\sqrt{\left( 2 \pi \right)^2\|V_z\|}} \mathrm{Exp} \left[ -\frac{1}{2} {\tilde{W}}^T V_z^{-1} {\tilde{W}} \right],
\end{equation}  
where 
\begin{equation}
\tilde{W} =  \mathrm{Log}_{z}(\tilde{X}) 
\end{equation}
represents the projection of the position $\tilde{X}$ from the Poincaré Disk into the tangent space $T_{z}\mathbb{D}$ centered on $z$, the hyperbolic barycenter of all occurrences for the set of points $X_i$. $\|V_z\|$ denotes the determinant of covariance matrix $V_z$ computed from the tangent space projection, and $j$ denotes the imaginary unit $j=\sqrt{-1}$.  

\section{Results}\label{Sec:results}
This section illustrates the effectiveness of the proposed hyperbolic embedding method for multilayer graphs through a series of case studies. We first apply the method to a multilayer stochastic block model (SBM), assessing both its capacity to recover underlying community structures and its robustness in scenarios where layers contain different sets of nodes. We then evaluate the approach on brain network data from patients with temporal lobe epilepsy, demonstrating its ability to yield consistent node localization across layers. Finally, we investigate the influence of the coupling parameter $\mu$ on the quality of the resulting embeddings. Together, these examples showcase the versatility and strengths of our method across diverse network analysis settings.

\subsection{Multilayer community structure – Multilayer SBM example}
To evaluate the capacity of our method to recover and emphasize community structures through hyperbolic embedding, we constructed a multilayer Stochastic Block Model (SBM). The model consists of \(n_l\) layers, each containing \(n\) nodes partitioned into \(n_c\) communities. Community assignments are kept consistent across layers. As in the classical single-layer SBM, nodes within the same community are more likely to be connected than nodes belonging to different communities. This intra-layer community structure is extended across layers to model inter-layer organization. 

Let us consider two nodes \(i\) and \(j\), assigned to communities \(c_i\) and \(c_j\), and belonging to layers \(l_i\) and \(l_j\), respectively. The probability \(p_{i,j}\) of an edge between these two nodes is defined as:

\begin{equation}
p_{i,j} = 
\begin{cases} 
p_{\neq} & \text{if } c_i \neq c_j \text{ and } l_i = l_j, \\
p_{=} & \text{if } c_i = c_j \text{ and } l_i = l_j, \\
\frac{p_{\neq}}{\alpha} & \text{if } c_i \neq c_j \text{ and } l_i \neq l_j, \\
\frac{p_{=}}{\alpha} & \text{if } c_i = c_j \text{ and } l_i \neq l_j. \\
\end{cases}
\end{equation}

%-------------Figure ------------
\begin{figure}[h!]
    \begin{center}
    \includegraphics[width=\columnwidth]{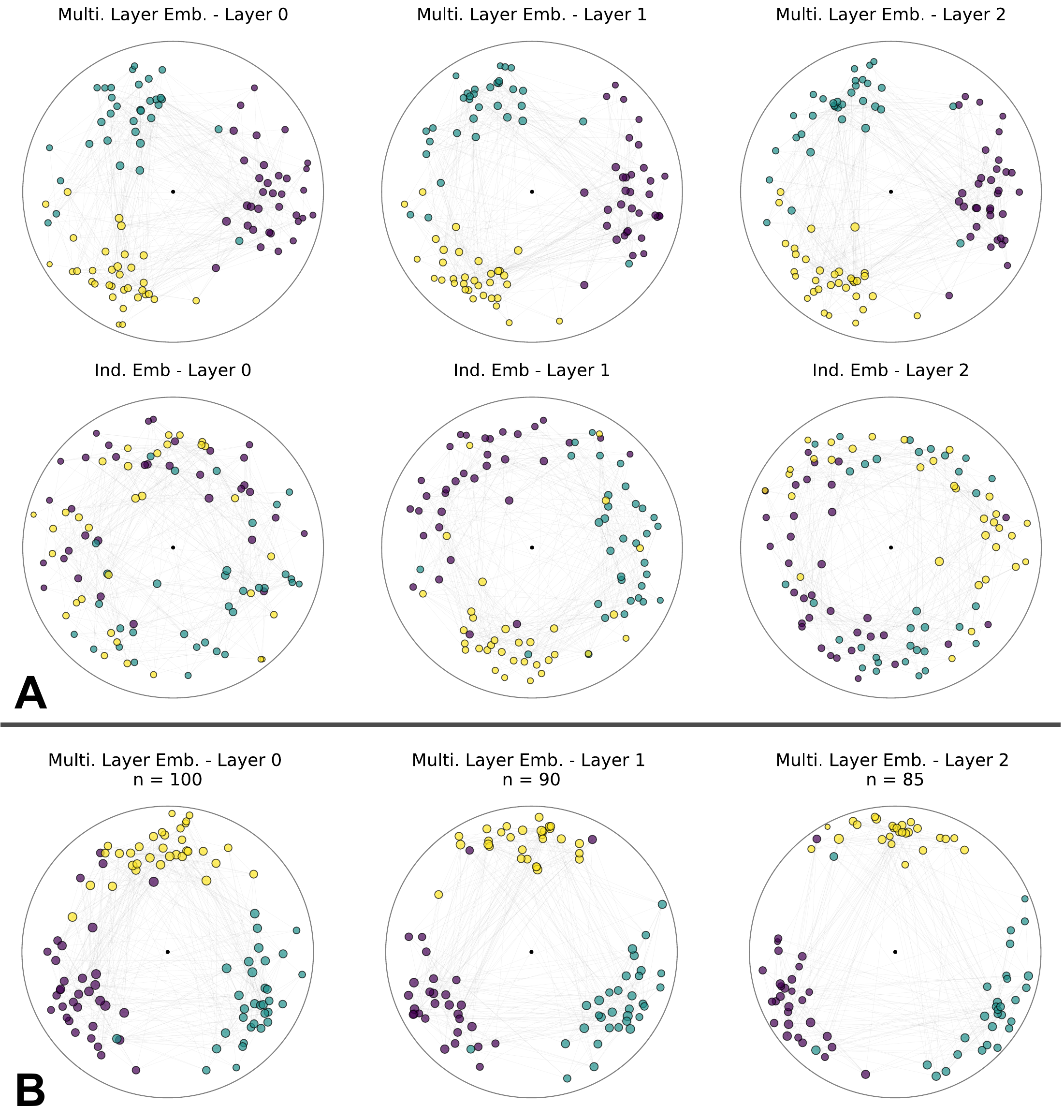}
    \caption{Hyperbolic embeddings of multilayer SBM models.
Parameters: 3 layers, 3 communities; $p_{=} = 0.16$, $p_{\neq} = 0.06$, $\mu = 20$, $\alpha = 10$. 
\textbf{(A)} Comparison between multilayer and independent embeddings for identical node sets across layers. 
Each layer contains 100 nodes. 
\textbf{Top row:} embeddings computed using the proposed multilayer approach. 
\textbf{Bottom row:} embeddings computed independently for each layer. 
\textbf{(B)} Multilayer embeddings with varying node sets across layers. 
The multilayer model comprises 3 layers with different numbers of nodes (Layer 0: 100 nodes; Layer 1: 90 nodes; Layer 2: 85 nodes).}
    \label{sbm_all}
    \end{center}
\end{figure}

Here, $p_{=}$ denotes the probability of a connection between two nodes belonging to the same community within the same layer, while $p_{\neq}$ corresponds to the probability of connexion for nodes in different communities within the same layer. To ensure a detectable community structure, we assume $p_{=} \geq p_{\neq}$. The parameter $\alpha \geq 1$ controls the strength of inter-layer coupling: $\alpha= 1$ corresponds to maximal coupling (no distinction between intra- and inter-layer edges), while larger values of $\alpha$ reduce inter-layer connectivity. 

To demonstrate the effectiveness of our approach in revealing community structure, we compared the results of the hyperbolic embedding in the Poincaré disk under two settings: \emph{i)} independent embedding of each layer without cross-layer information, and \emph{ii)} joint embedding of all layers using our multilayer method. Remarkably, even under weak inter-layer coupling (\(\alpha = 10\)), the multilayer approach reveals well-separated community clusters, unlike the independent embeddings where the structure appears more dispersed. These results, shown in Fig.~\ref{sbm_all}, highlight the advantage of the multilayer embedding in networks with latent community structure. When using our multilayer framework, the communities are clearly delineated in the hyperbolic space, whereas the independent approach requires stronger intra-community and lower inter-community connectivity to achieve comparable separation. \corr{This is due to the fact that a lower inter-community connectivity enhances the contrast between intra- and inter-community edge probabilities, which is essential for the independent embedding to produce a clear separation of communities. When this contrast diminishes and the connection probabilities become similar, the network approaches a regime akin to an Erdős–Rényi graph, where communities are less distinguishable without supplementary interlayer information}. \corr{When replacing Isomap with Laplacian eigenmaps on the same dataset, the multilayer embedding still preserves community separation, while the performance of the independent embedding further deteriorates with the spectral method.} 

Another key advantage of our method lies in its capacity to accommodate multilayer networks comprising different sets of nodes across layers, provided that a correspondence matrix between node-layer pairs is available. To assess this capability, we generated a three-layer stochastic block model (SBM) using our framework and randomly removed nodes from certain layers prior to embedding the network into the Poincaré disk. The results, depicted in Fig.~\ref{sbm_all}, illustrate that Layer 0 retains all 100 nodes, whereas Layers 1 and 2 contain only 90 and 85 nodes, respectively, due to the random removal of 10 and 15 nodes. Despite these discrepancies, our method effectively embeds all layers while preserving the latent community structure. These findings indicate that our approach robustly captures the multilayer community organization even when the individual layers consist of non-identical node sets.

\corr{Other multilayer/multiplex network models exist, including the Geometric Multiplex Model (GMM) \cite{vanderkolk2025multiplexity, Kleineberg2016}. We applied our embedding method to networks generated using the GMM model and compared the pairwise hyperbolic distances between nodes in the original configuration and the embedded version. Using a 4-layer network with 100 nodes, we observed a mean Spearman correlation of 0.79 between the original and embedded hyperbolic distances, indicating that the embedding effectively captures the network geometry of the network.”}

\subsection{Nodes localization – An example on brain epilepsy networks}
A potential application of our method lies in the comparison of network populations, such as comparing the brain networks of patients with those of healthy controls. The spatial localization of nodes in the hyperbolic disk could serve as a valuable biomarker. To illustrate this use case, we analyzed diffusion MRI data from patients with a left temporal lobe epilepsy ($n=19$) and healthy controls ($n=28$). Each brain network consists of 164 nodes, corresponding to regions defined by a standard anatomical atlas (Destrieux atlas). The dataset is described in detail in~\cite{BESSON2014}.  

To illustrate the practical utility of our approach, we computed the hyperbolic Gaussian distribution of node positions (specifically from the left temporal lobes) within the Poincaré disk (See Sec.~\ref{sec_gaussian_dis}). We evaluated two embedding strategies: our proposed multilayer method, in which each layer represents a different patient, and an independent embedding approach, wherein each patient is embedded separately and the resulting configurations are subsequently aligned via rotational transformation.\corrb{For the rotational transformation, reflection symmetry (axis inversion) is also tested. The alignment error is computed both with and without reflection, and the configuration with the lowest error is retained.}. \corr{The alignment is performed by rotating each embedding (beyond the reference layer) to minimize the sum of hyperbolic distances between corresponding nodes, effectively minimizing the $G_{\text{score}}$ defined Eq.~\ref{eq_g_Score}. This procedure, similar to the one described in \cite{LonghenaChaos2024}, ensures comparability across layers while preserving the geometric invariance of the embedding space.}

\begin{figure}[h!]
    \begin{center}
    \includegraphics[width=0.5\textwidth]{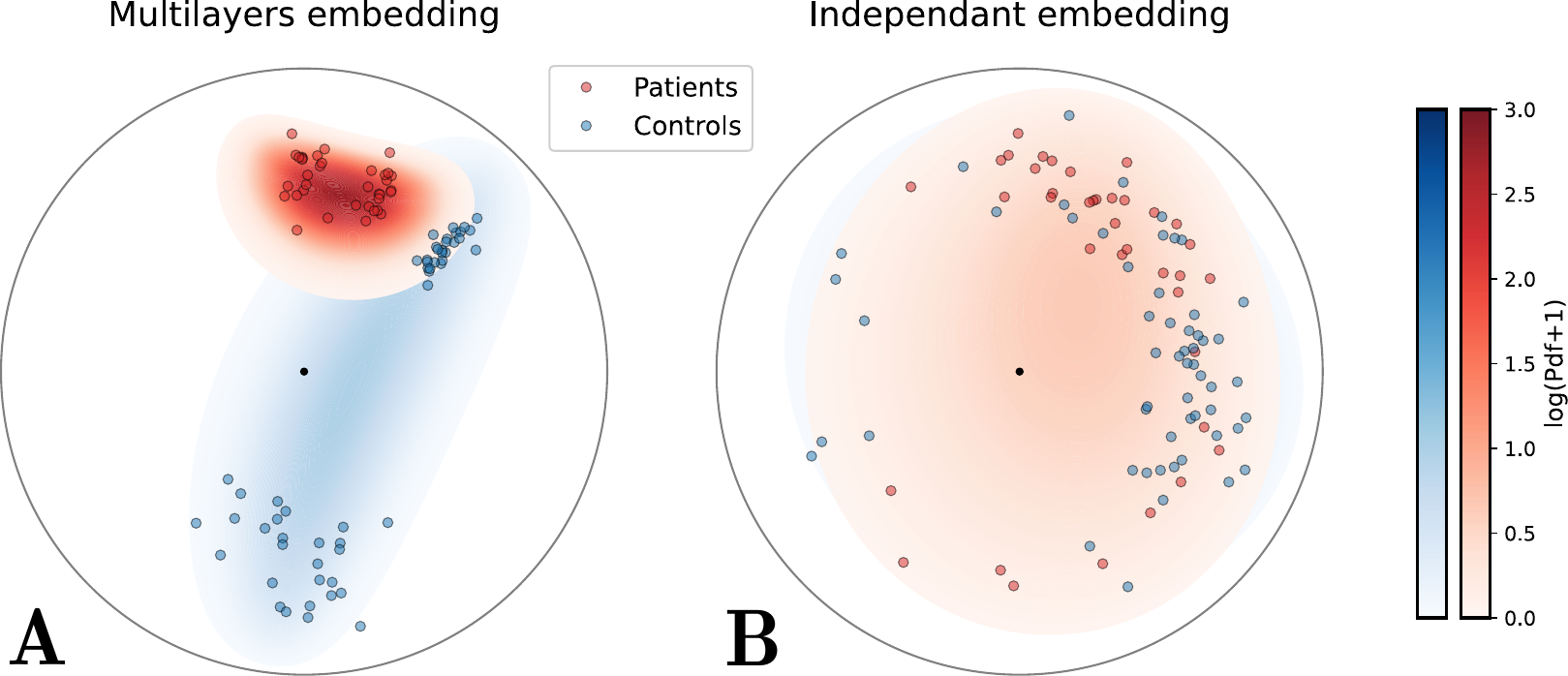}% distribution_epilesy_1.pdf}
    \caption{Hyperbolic embedding of patients with left temporal lobe epilepsy (red points) and controls (blue points). Visualization of the Gaussian distributions of node positions corresponding to the left temporal lobes \corr{(2 nodes per patients)} within the Poincaré disk, using two embedding strategies: A) the proposed multilayer embedding approach, where each layer represents an individual patient \corrb{$\beta = 50, \mu = 20$}; and B) an independent embedding of each patient followed by post hoc rotational alignment. }
    \label{epilepsy}
    \end{center}
\end{figure}

As shown in Fig.~\ref{epilepsy}, the multilayer embedding produces a more consistent and spatially coherent localization of the temporal lobe regions compared to the independent embedding strategy. From this result we observe that, compared with independently embedding each layer, the nodes within the left temporal lobe are more tightly clustered under our method, yielding a sharper Gaussian distribution. This enhanced localization highlights the capacity of the multilayer embedding to integrate information across patients, thereby generating more stable and reliable individual embeddings. Moreover, the method clearly uncovers distributional differences between patients with left temporal epilepsy and healthy controls, underscoring its potential utility in clinical settings. \corr{Permutation tests, where group labels were randomly reassigned, resulted in entropy distributions for both groups converging towards a common intermediate value, distinct from the lower entropy observed in patients (a more concentrated node positions) and higher entropy in controls (more dispersed configurations). This indicates that the tighter clustering in patients is not an artifact of the embedding method but reflects genuine group-specific network organization.}

%-------------Figure ------------
 
\subsection{Impact of the $\mu$ coupling parameter}
An important parameter in the method is the coupling strength $\mu$, which governs the degree of alignement between embeddings across layers. The effectiveness of this alignment can be quantitatively assessed using the global error score, defined as:
\begin{equation}
\label{eq_g_Score}
    G_{\text{score}}\left(E_1^h, E_2^h\right) = 
    \frac{1}{N} \sum_{i=1}^N d_{\mathbb{D}}\left(X_i^1, X_i^2\right) \, ,
\end{equation}
where $X_i^1$ and $X_i^2$ denote the positions of node $i$ in the first and second layer embeddings, respectively. $d_{\mathbb{D}}(\cdot, \cdot)$ is the hyperbolic distance function in the Poincaré's disk model (See Eq.~\ref{eq_hyp_dist}). $E_1^h$ and $E_2^h$ denote the embeddings of the first and second layer, respectively, in the Poincaré disk.
 In other words, the score corresponds to the mean hyperbolic displacement of nodes between the two layers. A lower $G_{\text{score}}$ indicates a better alignment between the embeddings, and thus a more coherent multilayer representation. 

\begin{figure}[h!]
    \begin{center}
    \begin{subfigure}[b]{0.5\textwidth}%0.8
        \centering
        \includegraphics[width=\textwidth]{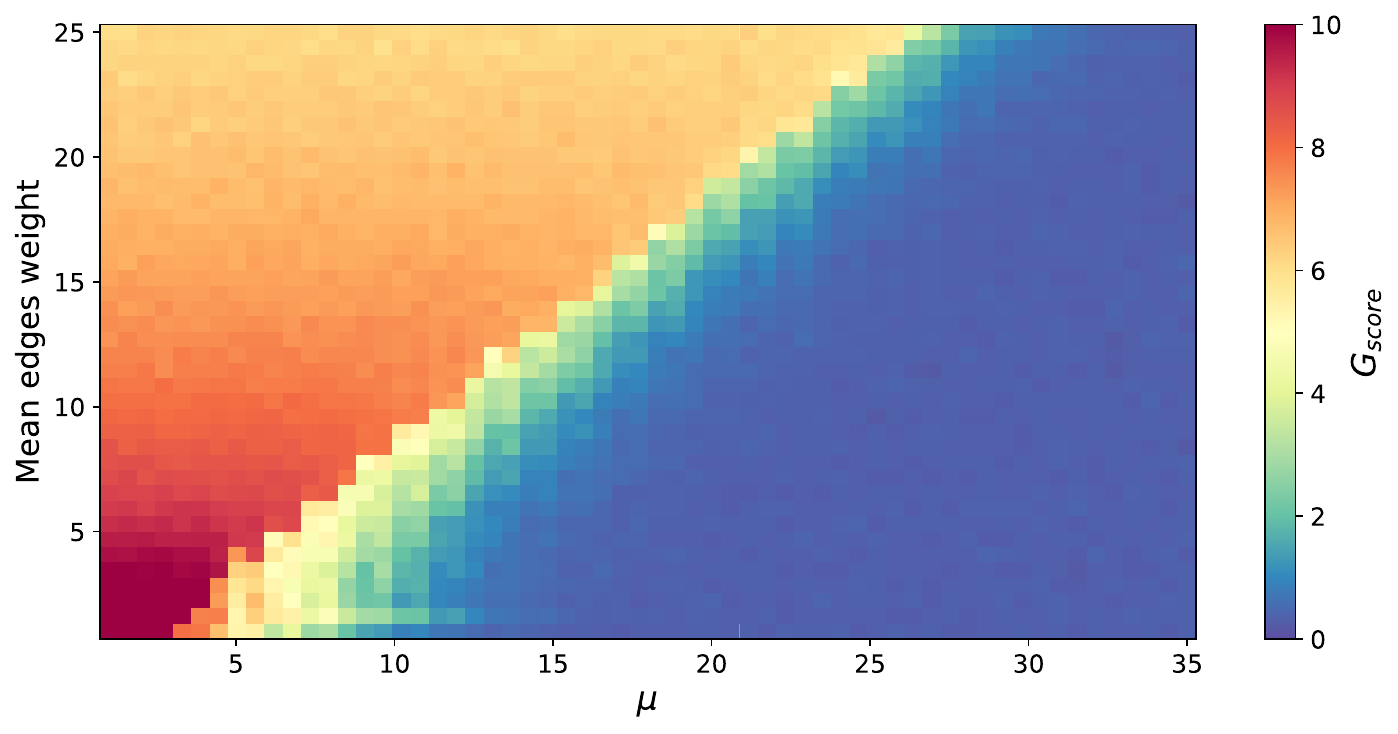}
        \caption{Two-layer SBM model with identical connectivity distributions.}
        \label{fig:heatmap_identical}
    \end{subfigure}
    \hfill
    \begin{subfigure}[b]{0.5\textwidth}%0.8
        \centering
        \includegraphics[width=\textwidth]{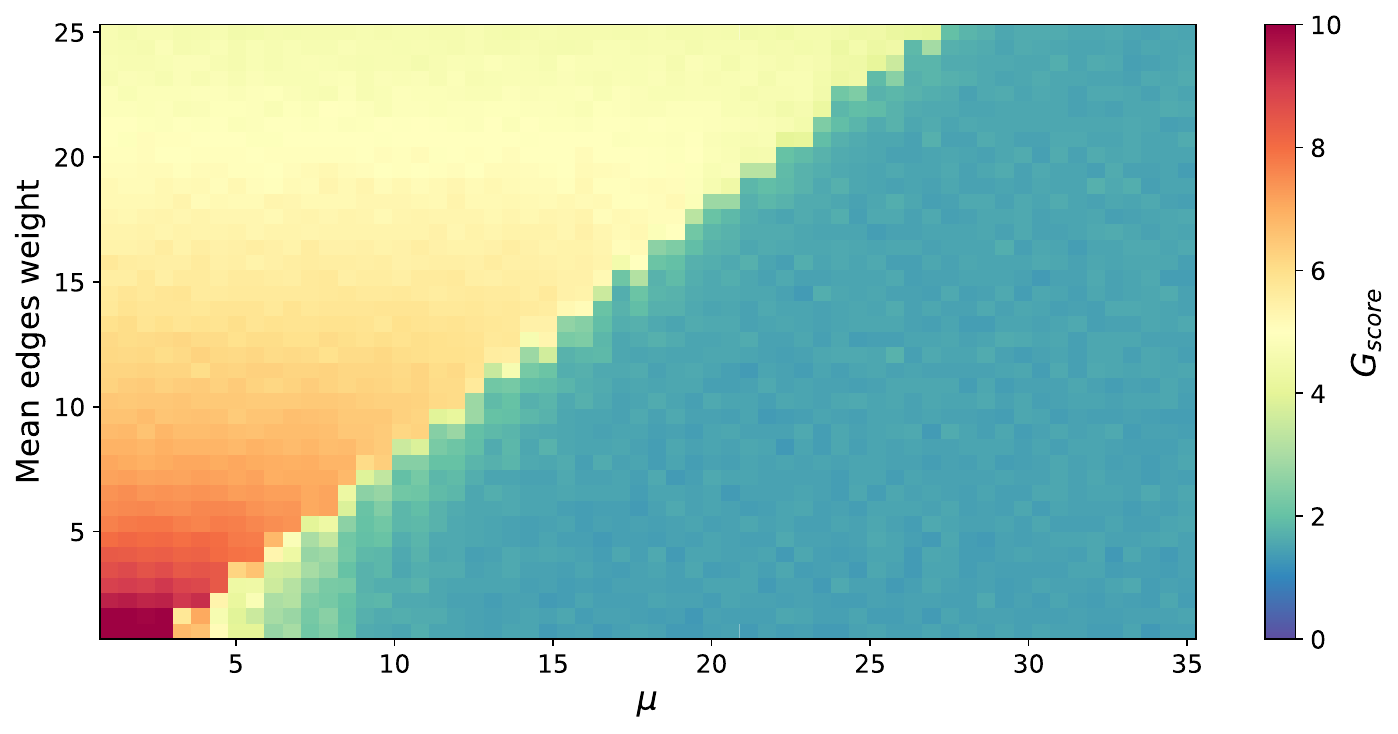}
        \caption{Two-layer SBM model with a different connectivity in the second layer.}
        \label{fig:heatmap_unidentical}
    \end{subfigure}
    \caption{$G_{\text{score}}$ (See Eq.~\ref{eq_g_Score}) as a function of the coupling parameter $\mu$ and the mean edge weight of the graph. Calculations were performed on a two-layer SBM model with $n = 100$ nodes per layer.}
    \label{fig:heatmap_mu}
    \end{center}
\end{figure}

To assess the impact of the coupling parameter $\mu$, we performed the embedding over a range of $\mu$ values for a two-layer graph (a multilayer SBM model), computing the corresponding $G_{\text{score}}$ in each case. Interestingly, we observed a sharp transition in the behavior of $G_{\text{score}}$ as $\mu$ increases. Beyond a critical value, denoted $\mu^*$, the score drops significantly and stabilizes at a plateau value, referred to as ${G_{\text{score}}}^*$ (see Fig.~\ref{fig:heatmap_mu}). Plateau is stable even when $\mu$ takes large values such as $\mu \to 10^4 \mu^*$ (result not shown). \corr{This indicates that, to reach a stable regime on the plateau, the parameter $\mu$ must satisfy $\mu \gg \mu^* \simeq$ the mean edge weight on the graph. Since the alignment remains stable once the plateau is reached, choosing a sufficiently high value of $\mu$ ensures convergence without adverse effects.}

To further interpret the physical meaning of $\mu^*$, we replicated this analysis using a multilayer stochastic block model (SBM) with varying the mean edge weights of the graphs. This was implemented by scaling the edge weights of the original graph to achieve the desired mean. Remarkably, we found that $\mu^*$ scales with the mean edge weight of the network, and the plateau value ${G_{\text{score}}}^*$ is consistently reached when $\mu$ matches this mean value. This suggests that optimal embedding performance occurs when the coupling strength $\mu$ reflects the average internal connectivity strength of the layers.

The global plateau value ${G_{\text{score}}}^*$ also appears to carry physical significance. While the critical coupling $\mu^*$ is governed by the average edge weight, the value of ${G_{\text{score}}}^*$ seems to reflect the similarity between the connectivity distributions of the layers. To assess this hypothesis, we perturbed the connection probabilities in the second layer (i.e., modifying $p_=$ and $p_{\neq}$) while keeping the first layer unchanged. As shown in Fig.~\ref{fig:heatmap_unidentical}, the relationship between $\mu^*$ and the mean edge weight remains robust to these perturbations, indicating that the phase transition is not disrupted. However, the plateau value ${G_{\text{score}}}^*$ increases when the two layers become less similar, compared to the case with identical probability distributions (Fig.~\ref{fig:heatmap_identical}). These results suggest that ${G_{\text{score}}}^*$ can serve as a quantitative indicator of interlayer similarity: the more similar the layers, the lower the score. \corr{To further validate this interpretation, we compared ${G_{\text{score}}}^*$ with the graph edit distance \cite{icpram15} in a controlled perturbation experiment, where edges were randomly added and removed from an initial network. The two metrics showed a strong linear relationship, with a Pearson correlation coefficient of 0.969, confirming that ${G_{\text{score}}}^*$ reliably captures structural dissimilarities between layers.}

%-------------Figure ------------

\section{Discussion}\label{Sec:discussion}
In this work, we have presented a novel framework for embedding multilayer networks into hyperbolic space by extending the coalescent embedding approach to incorporate inter-layer interactions via a global connectivity matrix, with the Poincaré disk serving as the embedding space. In contrast to prior methods such as~\cite{Kleineberg2016}, which embed each layer independently and perform comparative analyses post hoc, our approach integrates information from all layers simultaneously. This joint embedding strategy enhances the coherence and interpretability of the resulting representations, offering a significant advantage in contexts where structural patterns are shared across layers.

\corr{Our method is entirely data-driven and does not rely on explicit generative assumptions, in contrast to model-based approaches such as Mercator~\cite{Garcia2019}, which infer latent geometric parameters through a probabilistic framework grounded in network generative models. While Mercator provides robust parameter estimation and interpretable embeddings even in weakly geometric regimes, our approach offers complementary advantages by allowing flexible and direct embedding without requiring prior model parameter estimation. This makes our framework particularly suitable for empirical settings where model assumptions may not hold or where parameters are challenging to infer, such as in heterogeneous or partially observed multilayer networks.}

The embedding is performed in the Poincaré disk, a bounded model of hyperbolic space. This choice brings multiple benefits. First, it offers a well-defined geometric boundary at radius $r=1$, which naturally encodes disconnected or peripheral nodes addressing a common issue in the native hyperbolic disk where disconnected nodes tend toward infinity. Second, it allows for direct and intuitive comparison of node positions across layers, as all layers are mapped into a common finite geometric domain. Finally, it provides a visual interpretability, which are critical for downstream analyses such as classification or clustering.

A key strength of our method lies in its ability to embed layers with distinct node sets. Many real-world systems, particularly in biological and social contexts, exhibit partial or evolving node participation across layers. Our approach accommodates such heterogeneity by adapting the global connectivity matrix with rectangular coupling blocks that reflect known node correspondences. As demonstrated in the multilayer SBM experiments, our method successfully captures multilayer community structure even when layers do not share the same set of nodes,  highlighting its robustness and general applicability.

Compared to conventional approaches, where each layer is embedded independently and subsequently aligned, the proposed multilayer framework achieves more accurate spatial localization of nodes with similar connectivity across layers, as quantified through hyperbolic Gaussian distributions. When applied to real-world brain networks from patients with temporal lobe epilepsy, the method consistently identifies disease-related brain regions in the Poinacré Disk across individuals, outperforming the independent layer embedding approach and demonstrating its potential for comparative analysis in clinical neuroscience. These results indicate that our approach provides a principled basis for constructing group-representative networks by associating each node with a Gaussian probability distribution in the hyperbolic disk. The improved consistency of the embeddings suggests broader applicability to tasks such as anomaly detection, network-based phenotyping, and cross-subject analyses.

Another central contribution of this work is the analysis of the coupling parameter $\mu$, which modulates the influence of inter-layer interactions in the global embedding. This parameter governs the influence of inter-layer interactions in the global embedding. We show that the alignment quality of the embeddings undergoes a clear transition as $\mu$ increases: below a critical threshold $\mu^*$,  embeddings remain poorly aligned, whereas above $\mu^*$, a plateau is reached, indicating successful integration of layers.. Notably, we observe that $\mu^*$ scales with the average edge weight of the network, suggesting that optimal coupling reflects the intrinsic connectivity strength within the layers. Furthermore, the plateau value of the alignment score \corr{$G_{\mathrm{score}}$} serves as a proxy for inter-layer similarity, with lower scores corresponding to more structurally homogeneous layers. \corr{These findings suggest that the coupling parameter $\mu$ acts not only as a tunable parameter for controlling multilayer integration, but also as a diagnostic indicator of the underlying network structure. Specifically, the optimal value $\mu$ must be sufficiently large relative to the mean edge weight of the network to ensure stable alignment across layers. Furthermore, the plateau value of the alignment score, $G_{\text{score}}^*$, quantitatively reflects the degree of similarity or dissimilarity between layers, thus providing insight into the latent geometric and structural organization of the multilayer system.}  

In addition to structural analysis, the proposed framework offers new perspectives for perturbation detection. For instance, one can treat the original network as one layer and a perturbed version as another. The joint embedding then reveals how perturbations affect node positions in hyperbolic space. This approach provides a principled alternative to methods like~\cite{LonghenaChaos2024}, where networks are embedded separately and compared post-alignment. The inclusion of inter-layer coupling allows our method to capture both global and local deviations more effectively, offering a potential refinement to existing perturbation scoring techniques.

While this study focuses on two-dimensional embeddings for clarity and visualization, the framework is readily extendable to higher dimensions. Embedding into the $n$-dimensional Poincaré ball $\mathbb{D}^n$ would allow for the representation of more complex topological features and larger networks, in a similar spirit to the D-Mercator method~\cite{Jankowski2023}. Furthermore, our embedding pipeline, centered on the global connectivity matrix and the coupling parameter, can be generalized to other types of geometries or embedding techniques.

\corr{An important limitation of the proposed method concerns its computational scalability. The approach requires constructing and processing a global connectivity matrix of size $ M \times M $, where $ M = N \times L $, with $ N $ the number of nodes per layer and $ L $ the number of layers. Consequently, the computational complexity scales approximately as $ \mathcal{O}(M^2) $, which can quickly become prohibitive for multilayer networks with a large number of layers or nodes. This limitation restricts the applicability of our method to moderately sized multilayer systems, particularly in the context of temporal networks with high temporal resolution, where the number of layers (snapshots) is very large. Addressing this challenge will require algorithmic optimizations or approximation strategies, representing a promising direction for future work.}

In summary, this work provides a unified and flexible approach for embedding multilayer networks into hyperbolic space. It enhances interpretability, accommodates heterogeneity across layers, captures perturbations, and reveals meaningful geometric transitions through the coupling parameter. These contributions open new directions for the study of complex systems in a geometric framework.

\section*{Data Availability}
The connectivity data are not publicly available due to privacy and ethical restrictions. The code supporting the findings of this study are available at the following GitHub repository: \url{https://github.com/MartinG-38/MLNHypEmb}. 

\begin{acknowledgments}
M.G. acknowledges doctoral support from the Ecole Normale Supérieure de Lyon, France.\\
We thank Marián Ángeles Serrano for kindly sharing the implementation of the Geometric Multiplex Model (GMM), which was used in our simulations.

\end{acknowledgments}
%\appendix

\bibliography{biblio}% Produces the bibliography via BibTeX.

\begin{thebibliography}{38}%
\makeatletter
\providecommand \@ifxundefined [1]{%
 \@ifx{#1\undefined}
}%
\providecommand \@ifnum [1]{%
 \ifnum #1\expandafter \@firstoftwo
 \else \expandafter \@secondoftwo
 \fi
}%
\providecommand \@ifx [1]{%
 \ifx #1\expandafter \@firstoftwo
 \else \expandafter \@secondoftwo
 \fi
}%
\providecommand \natexlab [1]{#1}%
\providecommand \enquote  [1]{``#1''}%
\providecommand \bibnamefont  [1]{#1}%
\providecommand \bibfnamefont [1]{#1}%
\providecommand \citenamefont [1]{#1}%
\providecommand \href@noop [0]{\@secondoftwo}%
\providecommand \href [0]{\begingroup \@sanitize@url \@href}%
\providecommand \@href[1]{\@@startlink{#1}\@@href}%
\providecommand \@@href[1]{\endgroup#1\@@endlink}%
\providecommand \@sanitize@url [0]{\catcode `\\12\catcode `\$12\catcode `\&12\catcode `\#12\catcode `\^12\catcode `\_12\catcode `\%12\relax}%
\providecommand \@@startlink[1]{}%
\providecommand \@@endlink[0]{}%
\providecommand \url  [0]{\begingroup\@sanitize@url \@url }%
\providecommand \@url [1]{\endgroup\@href {#1}{\urlprefix }}%
\providecommand \urlprefix  [0]{URL }%
\providecommand \Eprint [0]{\href }%
\providecommand \doibase [0]{https://doi.org/}%
\providecommand \selectlanguage [0]{\@gobble}%
\providecommand \bibinfo  [0]{\@secondoftwo}%
\providecommand \bibfield  [0]{\@secondoftwo}%
\providecommand \translation [1]{[#1]}%
\providecommand \BibitemOpen [0]{}%
\providecommand \bibitemStop [0]{}%
\providecommand \bibitemNoStop [0]{.\EOS\space}%
\providecommand \EOS [0]{\spacefactor3000\relax}%
\providecommand \BibitemShut  [1]{\csname bibitem#1\endcsname}%
\let\auto@bib@innerbib\@empty
%</preamble>
\bibitem [{\citenamefont {Boccaletti}\ \emph {et~al.}(2006)\citenamefont {Boccaletti}, \citenamefont {Latora}, \citenamefont {Moreno}, \citenamefont {Chavez},\ and\ \citenamefont {Hwang}}]{Boccaletti2006}%
  \BibitemOpen
  \bibfield  {author} {\bibinfo {author} {\bibfnamefont {S.}~\bibnamefont {Boccaletti}}, \bibinfo {author} {\bibfnamefont {V.}~\bibnamefont {Latora}}, \bibinfo {author} {\bibfnamefont {Y.}~\bibnamefont {Moreno}}, \bibinfo {author} {\bibfnamefont {M.}~\bibnamefont {Chavez}},\ and\ \bibinfo {author} {\bibfnamefont {D.-U.}\ \bibnamefont {Hwang}},\ }\href {https://doi.org/10.1016/J.PHYSREP.2005.10.009} {\bibfield  {journal} {\bibinfo  {journal} {Physics Reports}\ }\textbf {\bibinfo {volume} {424}},\ \bibinfo {pages} {175} (\bibinfo {year} {2006})}\BibitemShut {NoStop}%
\bibitem [{\citenamefont {Boccaletti}\ \emph {et~al.}(2014)\citenamefont {Boccaletti}, \citenamefont {Bianconi}, \citenamefont {Criado}, \citenamefont {del Genio}, \citenamefont {{n}es}, \citenamefont {Romance}, \citenamefont {{n}a Nadal}, \citenamefont {Wang},\ and\ \citenamefont {Zanin}}]{Boccaletti2014}%
  \BibitemOpen
  \bibfield  {author} {\bibinfo {author} {\bibfnamefont {S.}~\bibnamefont {Boccaletti}}, \bibinfo {author} {\bibfnamefont {G.}~\bibnamefont {Bianconi}}, \bibinfo {author} {\bibfnamefont {R.}~\bibnamefont {Criado}}, \bibinfo {author} {\bibfnamefont {C.~I.}\ \bibnamefont {del Genio}}, \bibinfo {author} {\bibfnamefont {J.~G.-G.}\ \bibnamefont {{n}es}}, \bibinfo {author} {\bibfnamefont {M.}~\bibnamefont {Romance}}, \bibinfo {author} {\bibfnamefont {I.~S.}\ \bibnamefont {{n}a Nadal}}, \bibinfo {author} {\bibfnamefont {Z.}~\bibnamefont {Wang}},\ and\ \bibinfo {author} {\bibfnamefont {M.}~\bibnamefont {Zanin}},\ }\href {https://doi.org/10.1016/j.physrep.2014.07.001} {\bibfield  {journal} {\bibinfo  {journal} {Physics Reports}\ }\textbf {\bibinfo {volume} {544}},\ \bibinfo {pages} {1} (\bibinfo {year} {2014})}\BibitemShut {NoStop}%
\bibitem [{\citenamefont {Kivel\"a}\ \emph {et~al.}(2014)\citenamefont {Kivel\"a}, \citenamefont {Arenas}, \citenamefont {Barthelemy}, \citenamefont {Gleeson}, \citenamefont {Moreno},\ and\ \citenamefont {Porter}}]{Kivel2014}%
  \BibitemOpen
  \bibfield  {author} {\bibinfo {author} {\bibfnamefont {M.}~\bibnamefont {Kivel\"a}}, \bibinfo {author} {\bibfnamefont {A.}~\bibnamefont {Arenas}}, \bibinfo {author} {\bibfnamefont {M.}~\bibnamefont {Barthelemy}}, \bibinfo {author} {\bibfnamefont {J.~P.}\ \bibnamefont {Gleeson}}, \bibinfo {author} {\bibfnamefont {Y.}~\bibnamefont {Moreno}},\ and\ \bibinfo {author} {\bibfnamefont {M.~A.}\ \bibnamefont {Porter}},\ }\href {https://doi.org/10.1093/COMNET/CNU016} {\bibfield  {journal} {\bibinfo  {journal} {Journal of Complex Networks}\ }\textbf {\bibinfo {volume} {2}},\ \bibinfo {pages} {203} (\bibinfo {year} {2014})}\BibitemShut {NoStop}%
\bibitem [{\citenamefont {Xu}\ \emph {et~al.}(2020)\citenamefont {Xu}, \citenamefont {Nenning}, \citenamefont {Schwartz}, \citenamefont {Hong}, \citenamefont {Vogelstein}, \citenamefont {Goulas}, \citenamefont {Fair}, \citenamefont {Schroeder}, \citenamefont {Margulies}, \citenamefont {Smallwood}, \citenamefont {Milham},\ and\ \citenamefont {Langs}}]{Xu2020}%
  \BibitemOpen
  \bibfield  {author} {\bibinfo {author} {\bibfnamefont {T.}~\bibnamefont {Xu}}, \bibinfo {author} {\bibfnamefont {K.~H.}\ \bibnamefont {Nenning}}, \bibinfo {author} {\bibfnamefont {E.}~\bibnamefont {Schwartz}}, \bibinfo {author} {\bibfnamefont {S.~J.}\ \bibnamefont {Hong}}, \bibinfo {author} {\bibfnamefont {J.~T.}\ \bibnamefont {Vogelstein}}, \bibinfo {author} {\bibfnamefont {A.}~\bibnamefont {Goulas}}, \bibinfo {author} {\bibfnamefont {D.~A.}\ \bibnamefont {Fair}}, \bibinfo {author} {\bibfnamefont {C.~E.}\ \bibnamefont {Schroeder}}, \bibinfo {author} {\bibfnamefont {D.~S.}\ \bibnamefont {Margulies}}, \bibinfo {author} {\bibfnamefont {J.}~\bibnamefont {Smallwood}}, \bibinfo {author} {\bibfnamefont {M.~P.}\ \bibnamefont {Milham}},\ and\ \bibinfo {author} {\bibfnamefont {G.}~\bibnamefont {Langs}},\ }\href {https://doi.org/10.1016/J.NEUROIMAGE.2020.117346} {\bibfield  {journal} {\bibinfo  {journal} {NeuroImage}\ }\textbf {\bibinfo {volume} {223}},\ \bibinfo {pages} {117346} (\bibinfo {year} {2020})}\BibitemShut
  {NoStop}%
\bibitem [{\citenamefont {Puxeddu}\ \emph {et~al.}(2021)\citenamefont {Puxeddu}, \citenamefont {Petti},\ and\ \citenamefont {Astolfi}}]{Puxeddu2021}%
  \BibitemOpen
  \bibfield  {author} {\bibinfo {author} {\bibfnamefont {M.~G.}\ \bibnamefont {Puxeddu}}, \bibinfo {author} {\bibfnamefont {M.}~\bibnamefont {Petti}},\ and\ \bibinfo {author} {\bibfnamefont {L.}~\bibnamefont {Astolfi}},\ }\href {https://doi.org/10.3389/FNSYS.2021.624183/BIBTEX} {\bibfield  {journal} {\bibinfo  {journal} {Frontiers in Systems Neuroscience}\ }\textbf {\bibinfo {volume} {15}},\ \bibinfo {pages} {624183} (\bibinfo {year} {2021})}\BibitemShut {NoStop}%
\bibitem [{\citenamefont {Goyal}\ and\ \citenamefont {Ferrara}(2018)}]{Goyal2018}%
  \BibitemOpen
  \bibfield  {author} {\bibinfo {author} {\bibfnamefont {P.}~\bibnamefont {Goyal}}\ and\ \bibinfo {author} {\bibfnamefont {E.}~\bibnamefont {Ferrara}},\ }\href {https://doi.org/10.1016/J.KNOSYS.2018.03.022} {\bibfield  {journal} {\bibinfo  {journal} {Knowledge-Based Systems}\ }\textbf {\bibinfo {volume} {151}},\ \bibinfo {pages} {78} (\bibinfo {year} {2018})}\BibitemShut {NoStop}%
\bibitem [{\citenamefont {Cui}\ \emph {et~al.}(2019)\citenamefont {Cui}, \citenamefont {Wang}, \citenamefont {Pei},\ and\ \citenamefont {Zhu}}]{Cui2019}%
  \BibitemOpen
  \bibfield  {author} {\bibinfo {author} {\bibfnamefont {P.}~\bibnamefont {Cui}}, \bibinfo {author} {\bibfnamefont {X.}~\bibnamefont {Wang}}, \bibinfo {author} {\bibfnamefont {J.}~\bibnamefont {Pei}},\ and\ \bibinfo {author} {\bibfnamefont {W.}~\bibnamefont {Zhu}},\ }\href {https://doi.org/10.1109/TKDE.2018.2849727} {\bibfield  {journal} {\bibinfo  {journal} {IEEE Transactions on Knowledge and Data Engineering}\ }\textbf {\bibinfo {volume} {31}},\ \bibinfo {pages} {833} (\bibinfo {year} {2019})}\BibitemShut {NoStop}%
\bibitem [{\citenamefont {Xu}(2021)}]{xu2021understanding}%
  \BibitemOpen
  \bibfield  {author} {\bibinfo {author} {\bibfnamefont {M.}~\bibnamefont {Xu}},\ }\href {https://doi.org/10.1137/20M1386062} {\bibfield  {journal} {\bibinfo  {journal} {SIAM Review}\ }\textbf {\bibinfo {volume} {63}},\ \bibinfo {pages} {825} (\bibinfo {year} {2021})}\BibitemShut {NoStop}%
\bibitem [{\citenamefont {Liu}\ \emph {et~al.}(2020)\citenamefont {Liu}, \citenamefont {Huang}, \citenamefont {Yu}, \citenamefont {Fan},\ and\ \citenamefont {Dong}}]{Liu2020}%
  \BibitemOpen
  \bibfield  {author} {\bibinfo {author} {\bibfnamefont {Z.}~\bibnamefont {Liu}}, \bibinfo {author} {\bibfnamefont {C.}~\bibnamefont {Huang}}, \bibinfo {author} {\bibfnamefont {Y.}~\bibnamefont {Yu}}, \bibinfo {author} {\bibfnamefont {B.}~\bibnamefont {Fan}},\ and\ \bibinfo {author} {\bibfnamefont {J.}~\bibnamefont {Dong}},\ }in\ \href {https://doi.org/10.1145/3340531.3411944} {\emph {\bibinfo {booktitle} {International Conference on Information and Knowledge Management, Proceedings}}}\ (\bibinfo  {publisher} {Association for Computing Machinery},\ \bibinfo {year} {2020})\ pp.\ \bibinfo {pages} {995--1004}\BibitemShut {NoStop}%
\bibitem [{\citenamefont {Pio-Lopez}\ \emph {et~al.}(2021)\citenamefont {Pio-Lopez}, \citenamefont {Valdeolivas}, \citenamefont {Tichit}, \citenamefont {Remy},\ and\ \citenamefont {Baudot}}]{PioLopez2021}%
  \BibitemOpen
  \bibfield  {author} {\bibinfo {author} {\bibfnamefont {L.}~\bibnamefont {Pio-Lopez}}, \bibinfo {author} {\bibfnamefont {A.}~\bibnamefont {Valdeolivas}}, \bibinfo {author} {\bibfnamefont {L.}~\bibnamefont {Tichit}}, \bibinfo {author} {\bibfnamefont {{\'E}.}~\bibnamefont {Remy}},\ and\ \bibinfo {author} {\bibfnamefont {A.}~\bibnamefont {Baudot}},\ }\href {https://doi.org/10.1038/s41598-021-87987-1} {\bibfield  {journal} {\bibinfo  {journal} {Scientific reports}\ }\textbf {\bibinfo {volume} {11}},\ \bibinfo {pages} {8794} (\bibinfo {year} {2021})}\BibitemShut {NoStop}%
\bibitem [{\citenamefont {Liu}\ \emph {et~al.}(2017)\citenamefont {Liu}, \citenamefont {Chen}, \citenamefont {Yeung}, \citenamefont {Suzumura},\ and\ \citenamefont {Chen}}]{Liu2017}%
  \BibitemOpen
  \bibfield  {author} {\bibinfo {author} {\bibfnamefont {W.}~\bibnamefont {Liu}}, \bibinfo {author} {\bibfnamefont {P.~Y.}\ \bibnamefont {Chen}}, \bibinfo {author} {\bibfnamefont {S.}~\bibnamefont {Yeung}}, \bibinfo {author} {\bibfnamefont {T.}~\bibnamefont {Suzumura}},\ and\ \bibinfo {author} {\bibfnamefont {L.}~\bibnamefont {Chen}},\ }in\ \href {https://doi.org/10.1109/ICDMW.2017.23} {\emph {\bibinfo {booktitle} {IEEE International Conference on Data Mining Workshops, ICDMW}}},\ Vol.\ \bibinfo {volume} {2017-November}\ (\bibinfo  {publisher} {IEEE Computer Society},\ \bibinfo {year} {2017})\ pp.\ \bibinfo {pages} {134--141}\BibitemShut {NoStop}%
\bibitem [{\citenamefont {Wang}\ \emph {et~al.}(2023{\natexlab{a}})\citenamefont {Wang}, \citenamefont {Jiang}, \citenamefont {Jiang}, \citenamefont {Yi}, \citenamefont {Nie},\ and\ \citenamefont {Zhang}}]{Wang2023}%
  \BibitemOpen
  \bibfield  {author} {\bibinfo {author} {\bibfnamefont {Q.}~\bibnamefont {Wang}}, \bibinfo {author} {\bibfnamefont {H.}~\bibnamefont {Jiang}}, \bibinfo {author} {\bibfnamefont {Y.}~\bibnamefont {Jiang}}, \bibinfo {author} {\bibfnamefont {S.}~\bibnamefont {Yi}}, \bibinfo {author} {\bibfnamefont {Q.}~\bibnamefont {Nie}},\ and\ \bibinfo {author} {\bibfnamefont {G.}~\bibnamefont {Zhang}},\ }\href {https://doi.org/10.1016/J.DCAN.2022.10.002} {\bibfield  {journal} {\bibinfo  {journal} {Digital Communications and Networks}\ }\textbf {\bibinfo {volume} {9}},\ \bibinfo {pages} {1157} (\bibinfo {year} {2023}{\natexlab{a}})}\BibitemShut {NoStop}%
\bibitem [{\citenamefont {Park}\ \emph {et~al.}(2020)\citenamefont {Park}, \citenamefont {Han},\ and\ \citenamefont {Yu}}]{Park2020}%
  \BibitemOpen
  \bibfield  {author} {\bibinfo {author} {\bibfnamefont {C.}~\bibnamefont {Park}}, \bibinfo {author} {\bibfnamefont {J.}~\bibnamefont {Han}},\ and\ \bibinfo {author} {\bibfnamefont {H.}~\bibnamefont {Yu}},\ }\href {https://doi.org/10.1016/J.KNOSYS.2020.105861} {\bibfield  {journal} {\bibinfo  {journal} {Knowledge-Based Systems}\ }\textbf {\bibinfo {volume} {197}},\ \bibinfo {pages} {105861} (\bibinfo {year} {2020})}\BibitemShut {NoStop}%
\bibitem [{\citenamefont {Wang}\ \emph {et~al.}(2023{\natexlab{b}})\citenamefont {Wang}, \citenamefont {Chang}, \citenamefont {Fu},\ and\ \citenamefont {Zhao}}]{Wang2023b}%
  \BibitemOpen
  \bibfield  {author} {\bibinfo {author} {\bibfnamefont {Y.}~\bibnamefont {Wang}}, \bibinfo {author} {\bibfnamefont {D.}~\bibnamefont {Chang}}, \bibinfo {author} {\bibfnamefont {Z.}~\bibnamefont {Fu}},\ and\ \bibinfo {author} {\bibfnamefont {Y.}~\bibnamefont {Zhao}},\ }\href {https://doi.org/10.1109/TMM.2021.3136098} {\bibfield  {journal} {\bibinfo  {journal} {IEEE Transactions on Multimedia}\ }\textbf {\bibinfo {volume} {25}},\ \bibinfo {pages} {1008} (\bibinfo {year} {2023}{\natexlab{b}})}\BibitemShut {NoStop}%
\bibitem [{\citenamefont {Liu}\ \emph {et~al.}(2018)\citenamefont {Liu}, \citenamefont {He}, \citenamefont {Cao}, \citenamefont {Yu}, \citenamefont {Ragin},\ and\ \citenamefont {Leow}}]{Liu2018}%
  \BibitemOpen
  \bibfield  {author} {\bibinfo {author} {\bibfnamefont {Y.}~\bibnamefont {Liu}}, \bibinfo {author} {\bibfnamefont {L.}~\bibnamefont {He}}, \bibinfo {author} {\bibfnamefont {B.}~\bibnamefont {Cao}}, \bibinfo {author} {\bibfnamefont {P.~S.}\ \bibnamefont {Yu}}, \bibinfo {author} {\bibfnamefont {A.~B.}\ \bibnamefont {Ragin}},\ and\ \bibinfo {author} {\bibfnamefont {A.~D.}\ \bibnamefont {Leow}},\ }in\ \href {https://doi.org/10.1609/AAAI.V32I1.11288} {\emph {\bibinfo {booktitle} {Proceedings of the AAAI Conference on Artificial Intelligence}}},\ Vol.~\bibinfo {volume} {32}\ (\bibinfo  {publisher} {AAAI press},\ \bibinfo {year} {2018})\ pp.\ \bibinfo {pages} {117--124}\BibitemShut {NoStop}%
\bibitem [{\citenamefont {Li}\ \emph {et~al.}(2018)\citenamefont {Li}, \citenamefont {Chen}, \citenamefont {Tong},\ and\ \citenamefont {Liu}}]{Li2018}%
  \BibitemOpen
  \bibfield  {author} {\bibinfo {author} {\bibfnamefont {J.}~\bibnamefont {Li}}, \bibinfo {author} {\bibfnamefont {C.}~\bibnamefont {Chen}}, \bibinfo {author} {\bibfnamefont {H.}~\bibnamefont {Tong}},\ and\ \bibinfo {author} {\bibfnamefont {H.}~\bibnamefont {Liu}},\ }in\ \href {https://doi.org/10.1137/1.9781611975321.77} {\emph {\bibinfo {booktitle} {Proceedings of the 2018 SIAM International Conference on Data Mining}}}\ (\bibinfo  {publisher} {Society for Industrial and Applied Mathematics Publications},\ \bibinfo {year} {2018})\ pp.\ \bibinfo {pages} {684--692}\BibitemShut {NoStop}%
\bibitem [{\citenamefont {Boguñ\'a}\ \emph {et~al.}(2008)\citenamefont {Boguñ\'a}, \citenamefont {Krioukov},\ and\ \citenamefont {Claffy}}]{Bog2008}%
  \BibitemOpen
  \bibfield  {author} {\bibinfo {author} {\bibfnamefont {M.}~\bibnamefont {Boguñ\'a}}, \bibinfo {author} {\bibfnamefont {D.}~\bibnamefont {Krioukov}},\ and\ \bibinfo {author} {\bibfnamefont {K.~C.}\ \bibnamefont {Claffy}},\ }\href {https://doi.org/10.1038/nphys1130} {\bibfield  {journal} {\bibinfo  {journal} {Nature Physics 2009 5:1}\ }\textbf {\bibinfo {volume} {5}},\ \bibinfo {pages} {74} (\bibinfo {year} {2008})}\BibitemShut {NoStop}%
\bibitem [{\citenamefont {Krioukov}\ \emph {et~al.}(2010)\citenamefont {Krioukov}, \citenamefont {Papadopoulos}, \citenamefont {Kitsak}, \citenamefont {Vahdat},\ and\ \citenamefont {Boguñ\'a}}]{Krioukov2010}%
  \BibitemOpen
  \bibfield  {author} {\bibinfo {author} {\bibfnamefont {D.}~\bibnamefont {Krioukov}}, \bibinfo {author} {\bibfnamefont {F.}~\bibnamefont {Papadopoulos}}, \bibinfo {author} {\bibfnamefont {M.}~\bibnamefont {Kitsak}}, \bibinfo {author} {\bibfnamefont {A.}~\bibnamefont {Vahdat}},\ and\ \bibinfo {author} {\bibfnamefont {M.}~\bibnamefont {Boguñ\'a}},\ }\href {https://doi.org/10.1103/PHYSREVE.82.036106/FIGURES/13/MEDIUM} {\bibfield  {journal} {\bibinfo  {journal} {Physical Review E - Statistical, Nonlinear, and Soft Matter Physics}\ }\textbf {\bibinfo {volume} {82}},\ \bibinfo {pages} {036106} (\bibinfo {year} {2010})}\BibitemShut {NoStop}%
\bibitem [{\citenamefont {Boguñ\'a}\ \emph {et~al.}(2021)\citenamefont {Boguñ\'a}, \citenamefont {Bonamassa}, \citenamefont {Domenico}, \citenamefont {Havlin}, \citenamefont {Krioukov},\ and\ \citenamefont {Serrano}}]{Bogu2021}%
  \BibitemOpen
  \bibfield  {author} {\bibinfo {author} {\bibfnamefont {M.}~\bibnamefont {Boguñ\'a}}, \bibinfo {author} {\bibfnamefont {I.}~\bibnamefont {Bonamassa}}, \bibinfo {author} {\bibfnamefont {M.~D.}\ \bibnamefont {Domenico}}, \bibinfo {author} {\bibfnamefont {S.}~\bibnamefont {Havlin}}, \bibinfo {author} {\bibfnamefont {D.}~\bibnamefont {Krioukov}},\ and\ \bibinfo {author} {\bibfnamefont {M.~{\'A}.}\ \bibnamefont {Serrano}},\ }\href {https://doi.org/10.1038/s42254-020-00264-4} {\bibfield  {journal} {\bibinfo  {journal} {Nature Reviews Physics}\ }\textbf {\bibinfo {volume} {3}},\ \bibinfo {pages} {114} (\bibinfo {year} {2021})}\BibitemShut {NoStop}%
\bibitem [{\citenamefont {Nickel}\ and\ \citenamefont {Kiela}(2017)}]{nickel2017poincare}%
  \BibitemOpen
  \bibfield  {author} {\bibinfo {author} {\bibfnamefont {M.}~\bibnamefont {Nickel}}\ and\ \bibinfo {author} {\bibfnamefont {D.}~\bibnamefont {Kiela}},\ }in\ \href {https://proceedings.neurips.cc/paper_files/paper/2017/file/59dfa2df42d9e3d41f5b02bfc32229dd-Paper.pdf} {\emph {\bibinfo {booktitle} {Advances in Neural Information Processing Systems (NIPS 2017)}}},\ Vol.~\bibinfo {volume} {30},\ \bibinfo {editor} {edited by\ \bibinfo {editor} {\bibfnamefont {I.}~\bibnamefont {Guyon}}, \bibinfo {editor} {\bibfnamefont {U.~V.}\ \bibnamefont {Luxburg}}, \bibinfo {editor} {\bibfnamefont {S.}~\bibnamefont {Bengio}}, \bibinfo {editor} {\bibfnamefont {H.}~\bibnamefont {Wallach}}, \bibinfo {editor} {\bibfnamefont {R.}~\bibnamefont {Fergus}}, \bibinfo {editor} {\bibfnamefont {S.}~\bibnamefont {Vishwanathan}},\ and\ \bibinfo {editor} {\bibfnamefont {R.}~\bibnamefont {Garnett}}}\ (\bibinfo  {publisher} {Curran Associates, Inc.},\ \bibinfo {address} {Long Beach, CA, USA},\ \bibinfo {year} {2017})\ pp.\ \bibinfo {pages}
  {6341--6350}\BibitemShut {NoStop}%
\bibitem [{\citenamefont {Papadopoulos}\ \emph {et~al.}(2012)\citenamefont {Papadopoulos}, \citenamefont {Kitsak}, \citenamefont {Serrano}, \citenamefont {Boguñ\'a},\ and\ \citenamefont {Krioukov}}]{Papadopoulos2012}%
  \BibitemOpen
  \bibfield  {author} {\bibinfo {author} {\bibfnamefont {F.}~\bibnamefont {Papadopoulos}}, \bibinfo {author} {\bibfnamefont {M.}~\bibnamefont {Kitsak}}, \bibinfo {author} {\bibfnamefont {M.~{\'A}.}\ \bibnamefont {Serrano}}, \bibinfo {author} {\bibfnamefont {M.}~\bibnamefont {Boguñ\'a}},\ and\ \bibinfo {author} {\bibfnamefont {D.}~\bibnamefont {Krioukov}},\ }\href {https://doi.org/10.1038/nature11459} {\bibfield  {journal} {\bibinfo  {journal} {Nature 2012 489:7417}\ }\textbf {\bibinfo {volume} {489}},\ \bibinfo {pages} {537} (\bibinfo {year} {2012})}\BibitemShut {NoStop}%
\bibitem [{\citenamefont {Papadopoulos}\ \emph {et~al.}(2015)\citenamefont {Papadopoulos}, \citenamefont {Psomas},\ and\ \citenamefont {Krioukov}}]{Papadopoulos2015}%
  \BibitemOpen
  \bibfield  {author} {\bibinfo {author} {\bibfnamefont {F.}~\bibnamefont {Papadopoulos}}, \bibinfo {author} {\bibfnamefont {C.}~\bibnamefont {Psomas}},\ and\ \bibinfo {author} {\bibfnamefont {D.}~\bibnamefont {Krioukov}},\ }\href {https://doi.org/10.1109/TNET.2013.2294052} {\bibfield  {journal} {\bibinfo  {journal} {IEEE/ACM Transactions on Networking}\ }\textbf {\bibinfo {volume} {23}},\ \bibinfo {pages} {198} (\bibinfo {year} {2015})}\BibitemShut {NoStop}%
\bibitem [{\citenamefont {Garc\'{\i}a-P\'erez}\ \emph {et~al.}(2019)\citenamefont {Garc\'{\i}a-P\'erez}, \citenamefont {Allard}, \citenamefont {Serrano},\ and\ \citenamefont {Boguñ\'a}}]{Garcia2019}%
  \BibitemOpen
  \bibfield  {author} {\bibinfo {author} {\bibfnamefont {G.}~\bibnamefont {Garc\'{\i}a-P\'erez}}, \bibinfo {author} {\bibfnamefont {A.}~\bibnamefont {Allard}}, \bibinfo {author} {\bibfnamefont {M.~{\'A}.}\ \bibnamefont {Serrano}},\ and\ \bibinfo {author} {\bibfnamefont {M.}~\bibnamefont {Boguñ\'a}},\ }\href {https://doi.org/10.1088/1367-2630/AB57D2} {\bibfield  {journal} {\bibinfo  {journal} {New Journal of Physics}\ }\textbf {\bibinfo {volume} {21}},\ \bibinfo {pages} {123033} (\bibinfo {year} {2019})}\BibitemShut {NoStop}%
\bibitem [{\citenamefont {Jankowski}\ \emph {et~al.}(2023)\citenamefont {Jankowski}, \citenamefont {Allard}, \citenamefont {Boguñ\'a},\ and\ \citenamefont {Serrano}}]{Jankowski2023}%
  \BibitemOpen
  \bibfield  {author} {\bibinfo {author} {\bibfnamefont {R.}~\bibnamefont {Jankowski}}, \bibinfo {author} {\bibfnamefont {A.}~\bibnamefont {Allard}}, \bibinfo {author} {\bibfnamefont {M.}~\bibnamefont {Boguñ\'a}},\ and\ \bibinfo {author} {\bibfnamefont {M.~{\'A}.}\ \bibnamefont {Serrano}},\ }\href {https://doi.org/10.1038/s41467-023-43337-5} {\bibfield  {journal} {\bibinfo  {journal} {Nature Communications 2023 14:1}\ }\textbf {\bibinfo {volume} {14}},\ \bibinfo {pages} {1} (\bibinfo {year} {2023})}\BibitemShut {NoStop}%
\bibitem [{\citenamefont {Muscoloni}\ \emph {et~al.}(2017)\citenamefont {Muscoloni}, \citenamefont {Thomas}, \citenamefont {Ciucci}, \citenamefont {Bianconi},\ and\ \citenamefont {Cannistraci}}]{Muscoloni2017}%
  \BibitemOpen
  \bibfield  {author} {\bibinfo {author} {\bibfnamefont {A.}~\bibnamefont {Muscoloni}}, \bibinfo {author} {\bibfnamefont {J.~M.}\ \bibnamefont {Thomas}}, \bibinfo {author} {\bibfnamefont {S.}~\bibnamefont {Ciucci}}, \bibinfo {author} {\bibfnamefont {G.}~\bibnamefont {Bianconi}},\ and\ \bibinfo {author} {\bibfnamefont {C.~V.}\ \bibnamefont {Cannistraci}},\ }\href {https://doi.org/10.1038/s41467-017-01825-5} {\bibfield  {journal} {\bibinfo  {journal} {Nature Communications 2017 8:1}\ }\textbf {\bibinfo {volume} {8}},\ \bibinfo {pages} {1} (\bibinfo {year} {2017})}\BibitemShut {NoStop}%
\bibitem [{\citenamefont {Kovács}\ and\ \citenamefont {Palla}(2021)}]{Kovcs2021}%
  \BibitemOpen
  \bibfield  {author} {\bibinfo {author} {\bibfnamefont {B.}~\bibnamefont {Kovács}}\ and\ \bibinfo {author} {\bibfnamefont {G.}~\bibnamefont {Palla}},\ }\href {https://doi.org/10.1038/s41598-021-87333-5} {\bibfield  {journal} {\bibinfo  {journal} {Scientific Reports 2021 11:1}\ }\textbf {\bibinfo {volume} {11}},\ \bibinfo {pages} {1} (\bibinfo {year} {2021})}\BibitemShut {NoStop}%
\bibitem [{\citenamefont {Wang}\ \emph {et~al.}(2021)\citenamefont {Wang}, \citenamefont {Huang}, \citenamefont {Ma}, \citenamefont {Liu},\ and\ \citenamefont {Vosoughi}}]{Wang2021}%
  \BibitemOpen
  \bibfield  {author} {\bibinfo {author} {\bibfnamefont {L.}~\bibnamefont {Wang}}, \bibinfo {author} {\bibfnamefont {C.}~\bibnamefont {Huang}}, \bibinfo {author} {\bibfnamefont {W.}~\bibnamefont {Ma}}, \bibinfo {author} {\bibfnamefont {R.}~\bibnamefont {Liu}},\ and\ \bibinfo {author} {\bibfnamefont {S.}~\bibnamefont {Vosoughi}},\ }\href {https://doi.org/10.1007/S10618-021-00774-4/TABLES/8} {\bibfield  {journal} {\bibinfo  {journal} {Data Mining and Knowledge Discovery}\ }\textbf {\bibinfo {volume} {35}},\ \bibinfo {pages} {1906} (\bibinfo {year} {2021})}\BibitemShut {NoStop}%
\bibitem [{\citenamefont {Yang}\ \emph {et~al.}(2023)\citenamefont {Yang}, \citenamefont {Zhou}, \citenamefont {Xiong},\ and\ \citenamefont {King}}]{Yang2023}%
  \BibitemOpen
  \bibfield  {author} {\bibinfo {author} {\bibfnamefont {M.}~\bibnamefont {Yang}}, \bibinfo {author} {\bibfnamefont {M.}~\bibnamefont {Zhou}}, \bibinfo {author} {\bibfnamefont {H.}~\bibnamefont {Xiong}},\ and\ \bibinfo {author} {\bibfnamefont {I.}~\bibnamefont {King}},\ }\href {https://doi.org/10.1109/TKDE.2022.3232398} {\bibfield  {journal} {\bibinfo  {journal} {IEEE Transactions on Knowledge and Data Engineering}\ }\textbf {\bibinfo {volume} {35}},\ \bibinfo {pages} {11489} (\bibinfo {year} {2023})}\BibitemShut {NoStop}%
\bibitem [{\citenamefont {Kleineberg}\ \emph {et~al.}(2016)\citenamefont {Kleineberg}, \citenamefont {Boguñ\'a}, \citenamefont {Serrano},\ and\ \citenamefont {Papadopoulos}}]{Kleineberg2016}%
  \BibitemOpen
  \bibfield  {author} {\bibinfo {author} {\bibfnamefont {K.~K.}\ \bibnamefont {Kleineberg}}, \bibinfo {author} {\bibfnamefont {M.}~\bibnamefont {Boguñ\'a}}, \bibinfo {author} {\bibfnamefont {M.~{\'A}.}\ \bibnamefont {Serrano}},\ and\ \bibinfo {author} {\bibfnamefont {F.}~\bibnamefont {Papadopoulos}},\ }\href {https://doi.org/10.1038/nphys3812} {\bibfield  {journal} {\bibinfo  {journal} {Nature Physics 2016 12:11}\ }\textbf {\bibinfo {volume} {12}},\ \bibinfo {pages} {1076} (\bibinfo {year} {2016})}\BibitemShut {NoStop}%
\bibitem [{\citenamefont {Von~Luxburg}(2007)}]{von2007tutorial}%
  \BibitemOpen
  \bibfield  {author} {\bibinfo {author} {\bibfnamefont {U.}~\bibnamefont {Von~Luxburg}},\ }\href {https://doi.org/10.1007/s11222-007-9033-z} {\bibfield  {journal} {\bibinfo  {journal} {Statistics and computing}\ }\textbf {\bibinfo {volume} {17}},\ \bibinfo {pages} {395} (\bibinfo {year} {2007})}\BibitemShut {NoStop}%
\bibitem [{\citenamefont {Tenenbaum}\ \emph {et~al.}(2000)\citenamefont {Tenenbaum}, \citenamefont {de~Silva},\ and\ \citenamefont {Langford}}]{Tenenbaum2000}%
  \BibitemOpen
  \bibfield  {author} {\bibinfo {author} {\bibfnamefont {J.~B.}\ \bibnamefont {Tenenbaum}}, \bibinfo {author} {\bibfnamefont {V.}~\bibnamefont {de~Silva}},\ and\ \bibinfo {author} {\bibfnamefont {J.~C.}\ \bibnamefont {Langford}},\ }\href {https://doi.org/10.1126/science.290.5500.2319} {\bibfield  {journal} {\bibinfo  {journal} {Science}\ }\textbf {\bibinfo {volume} {290}},\ \bibinfo {pages} {2319} (\bibinfo {year} {2000})}\BibitemShut {NoStop}%
\bibitem [{\citenamefont {Battiston}\ \emph {et~al.}(2014)\citenamefont {Battiston}, \citenamefont {Nicosia},\ and\ \citenamefont {Latora}}]{battiston2014structural}%
  \BibitemOpen
  \bibfield  {author} {\bibinfo {author} {\bibfnamefont {F.}~\bibnamefont {Battiston}}, \bibinfo {author} {\bibfnamefont {V.}~\bibnamefont {Nicosia}},\ and\ \bibinfo {author} {\bibfnamefont {V.}~\bibnamefont {Latora}},\ }\href@noop {} {\bibfield  {journal} {\bibinfo  {journal} {Physical Review E}\ }\textbf {\bibinfo {volume} {89}},\ \bibinfo {pages} {032804} (\bibinfo {year} {2014})}\BibitemShut {NoStop}%
\bibitem [{\citenamefont {Khrulkov}\ \emph {et~al.}(2020)\citenamefont {Khrulkov}, \citenamefont {Mirvakhabova}, \citenamefont {Ustinova}, \citenamefont {Oseledets},\ and\ \citenamefont {Lempitsky}}]{Khrulkov2020}%
  \BibitemOpen
  \bibfield  {author} {\bibinfo {author} {\bibfnamefont {V.}~\bibnamefont {Khrulkov}}, \bibinfo {author} {\bibfnamefont {L.}~\bibnamefont {Mirvakhabova}}, \bibinfo {author} {\bibfnamefont {E.}~\bibnamefont {Ustinova}}, \bibinfo {author} {\bibfnamefont {I.}~\bibnamefont {Oseledets}},\ and\ \bibinfo {author} {\bibfnamefont {V.}~\bibnamefont {Lempitsky}},\ }in\ \href {https://doi.org/10.1109/CVPR42600.2020.00645} {\emph {\bibinfo {booktitle} {2020 IEEE/CVF Conference on Computer Vision and Pattern Recognition (CVPR)}}}\ (\bibinfo {year} {2020})\ pp.\ \bibinfo {pages} {6418--6428}\BibitemShut {NoStop}%
\bibitem [{\citenamefont {Ungar}(2001)}]{ungar2001hyperbolic}%
  \BibitemOpen
  \bibfield  {author} {\bibinfo {author} {\bibfnamefont {A.~A.}\ \bibnamefont {Ungar}},\ }\href@noop {} {\bibfield  {journal} {\bibinfo  {journal} {Computers \& Mathematics with Applications}\ }\textbf {\bibinfo {volume} {41}},\ \bibinfo {pages} {135} (\bibinfo {year} {2001})}\BibitemShut {NoStop}%
\bibitem [{\citenamefont {van~der Kolk}\ \emph {et~al.}(2025)\citenamefont {van~der Kolk}, \citenamefont {Krioukov}, \citenamefont {Boguñá},\ and\ \citenamefont {Ángeles Serrano}}]{vanderkolk2025multiplexity}%
  \BibitemOpen
  \bibfield  {author} {\bibinfo {author} {\bibfnamefont {J.}~\bibnamefont {van~der Kolk}}, \bibinfo {author} {\bibfnamefont {D.}~\bibnamefont {Krioukov}}, \bibinfo {author} {\bibfnamefont {M.}~\bibnamefont {Boguñá}},\ and\ \bibinfo {author} {\bibfnamefont {M.}~\bibnamefont {Ángeles Serrano}},\ }\href {https://doi.org/10.48550/arXiv.2505.17688} {\bibinfo {title} {Multiplexity amplifies geometry in networks}} (\bibinfo {year} {2025}),\ \Eprint {https://arxiv.org/abs/2505.17688} {arXiv:2505.17688 [physics.soc-ph]} \BibitemShut {NoStop}%
\bibitem [{\citenamefont {Besson}\ \emph {et~al.}(2014)\citenamefont {Besson}, \citenamefont {Dinkelacker}, \citenamefont {Valabregue}, \citenamefont {Thivard}, \citenamefont {Leclerc}, \citenamefont {Baulac}, \citenamefont {Sammler}, \citenamefont {Colliot}, \citenamefont {Leh\'ericy}, \citenamefont {Samson},\ and\ \citenamefont {Dupont}}]{BESSON2014}%
  \BibitemOpen
  \bibfield  {author} {\bibinfo {author} {\bibfnamefont {P.}~\bibnamefont {Besson}}, \bibinfo {author} {\bibfnamefont {V.}~\bibnamefont {Dinkelacker}}, \bibinfo {author} {\bibfnamefont {R.}~\bibnamefont {Valabregue}}, \bibinfo {author} {\bibfnamefont {L.}~\bibnamefont {Thivard}}, \bibinfo {author} {\bibfnamefont {X.}~\bibnamefont {Leclerc}}, \bibinfo {author} {\bibfnamefont {M.}~\bibnamefont {Baulac}}, \bibinfo {author} {\bibfnamefont {D.}~\bibnamefont {Sammler}}, \bibinfo {author} {\bibfnamefont {O.}~\bibnamefont {Colliot}}, \bibinfo {author} {\bibfnamefont {S.}~\bibnamefont {Leh\'ericy}}, \bibinfo {author} {\bibfnamefont {S.}~\bibnamefont {Samson}},\ and\ \bibinfo {author} {\bibfnamefont {S.}~\bibnamefont {Dupont}},\ }\href {https://doi.org/https://doi.org/10.1016/j.neuroimage.2014.04.071} {\bibfield  {journal} {\bibinfo  {journal} {NeuroImage}\ }\textbf {\bibinfo {volume} {100}},\ \bibinfo {pages} {135} (\bibinfo {year} {2014})}\BibitemShut {NoStop}%
\bibitem [{\citenamefont {Longhena}\ \emph {et~al.}(2024)\citenamefont {Longhena}, \citenamefont {Guillemaud},\ and\ \citenamefont {Chavez}}]{LonghenaChaos2024}%
  \BibitemOpen
  \bibfield  {author} {\bibinfo {author} {\bibfnamefont {A.}~\bibnamefont {Longhena}}, \bibinfo {author} {\bibfnamefont {M.}~\bibnamefont {Guillemaud}},\ and\ \bibinfo {author} {\bibfnamefont {M.}~\bibnamefont {Chavez}},\ }\href {https://doi.org/10.1063/5.0199546} {\bibfield  {journal} {\bibinfo  {journal} {Chaos: An Interdisciplinary Journal of Nonlinear Science}\ }\textbf {\bibinfo {volume} {34}},\ \bibinfo {pages} {063117} (\bibinfo {year} {2024})}\BibitemShut {NoStop}%
\bibitem [{\citenamefont {Abu{-}Aisheh}\ \emph {et~al.}(2015)\citenamefont {Abu{-}Aisheh}, \citenamefont {Raveaux}, \citenamefont {Ramel},\ and\ \citenamefont {Martineau}}]{icpram15}%
  \BibitemOpen
  \bibfield  {author} {\bibinfo {author} {\bibfnamefont {Z.}~\bibnamefont {Abu{-}Aisheh}}, \bibinfo {author} {\bibfnamefont {R.}~\bibnamefont {Raveaux}}, \bibinfo {author} {\bibfnamefont {J.}~\bibnamefont {Ramel}},\ and\ \bibinfo {author} {\bibfnamefont {P.}~\bibnamefont {Martineau}},\ }in\ \href {https://doi.org/10.5220/0005209202710278} {\emph {\bibinfo {booktitle} {Proceedings of the International Conference on Pattern Recognition Applications and Methods - Volume 1: ICPRAM}}},\ \bibinfo {organization} {INSTICC}\ (\bibinfo  {publisher} {SciTePress},\ \bibinfo {year} {2015})\ pp.\ \bibinfo {pages} {271--278}\BibitemShut {NoStop}%
\end{thebibliography}%
\end{document}